\RequirePackage{fix-cm}
\documentclass[twocolumn,epjc3]{svjour3} 
\smartqed  
\RequirePackage{graphicx,hyphenat}
\journalname{Eur. Phys. J. C}

\usepackage{graphicx,color,soul}
\usepackage{hyphenat}
\def\be{\begin{eqnarray} &&}
\def\nonu{\nonumber \\ &&}

\def\ee{\end{eqnarray}}
\def\psla{\slash \! \! \!}
\newcommand{\eqref}[1]{(\ref{#1})}

\begin{document}
\title{ Fermionic bound states in Minkowski-space: \\   Light-cone singularities and structure
}
\author{Wayne de Paula \thanksref{e0,addr1}\and Tobias Frederico 
\thanksref{e1,addr1}\and Giovanni Salm\`e
\thanksref{e2,addr2}\and
Michele Viviani \thanksref{e3,addr3} \and Rafael Pimentel \thanksref{e4,addr1}}

\thankstext{e0}{e-mail: wayne@ita.br}
\thankstext{e1}{e-mail: tobias@ita.br}
\thankstext{e2}{e-mail: salmeg@roma1.infn.it}
\thankstext{e3}{e-mail: michele.viviani@pi.infn.it}
\thankstext{e4}{e-mail: pimentel.es@gmail.com}

\institute{Dep. de F\'\i sica, Instituto Tecnol\'ogico de Aeron\'autica,
DCTA, 12.228-900 S\~ao Jos\'e dos
Campos, S\~ao Paulo, Brazil \label{addr1}\and Istituto  Nazionale di Fisica Nucleare, Sezione di Roma, P.le A. Moro 2,
 I-00185 Roma, Italy \label{addr2}\and Istituto  Nazionale di Fisica Nucleare, Sezione di Pisa,
Largo Pontecorvo 3, 56100, Pisa, Italy \label{addr3} 
}

\date{Received: date / Accepted: date}

\maketitle

\begin{abstract}
The Bethe-Salpeter equation for two-body bound system with  spin
$1/2$ constituent is addressed directly in the Minkowski space. In order to
accomplish this aim we use the Nakanishi integral representation of the
Bethe-Salpeter amplitude and exploit the formal tool represented by the exact 
  projection onto the null-plane. This formal step allows one i) to deal with 
 end-point singularities one meets and ii) to find stable results, up to
 strongly relativistic regimes, { that settles in strongly bound systems}.
  We apply this technique to obtain { the numerical dependence of the binding
  energies upon the coupling constants and the light-front amplitudes}  for
 a fermion-fermion $0^+$ state with interaction kernels, in ladder approximation, corresponding 
 to scalar-,
 pseudoscalar- and vector\hyp{}boson exchanges, respectively. After completing the numerical survey
 of the previous cases, we extend our approach to a quark-antiquark system
 in $0^-$ state, taking both  constituent-fermion 
 and exchanged\hyp{}boson masses, from lattice calculations.
 Interestingly,    the calculated light-front
 amplitudes for such a {\it mock pion} show peculiar signatures of the spin degrees of freedom.

\end{abstract}
\keywords{Bethe-Salpeter equation,  Minkowski space, integral
representation, 
Light-front projection, fermion bound states }
\maketitle
\flushbottom

\section{Introduction}
The standard approach to the relativistic bound state problem in quantum field theory was
formulated, more than a half century ago, in a seminal work by Salpeter and Bethe \cite{BS51}. 
{  In principle, the Bethe-Salpeter equation (BSE) 
 (see also the review  \cite{Nakarev}) 
allows one 
to access the   non perturbative regime of  the dynamics inside a relativistic interacting system, 
as the
Schr\"odinger equation does in a non relativistic regime.} 
As it is well-known, 
apart the celebrated Wick-Cutkosky model
 \cite{wick1954properties,cutkosky1954solutions},
 composed by two scalars exchanging a massless scalar, solving BSE
  { is  very difficult when one adopts the variables of the  space  where
  the physical processes take place, namely the Minkowski space.} 
Furthermore  
 the irreducible kernel itself cannot be written in a closed
form. Nonetheless, 
in hadron physics,
it could be highly desirable to  develop non perturbative tools in Minkowski space 
suitable for  supporting, { e.g.,} 
 experimental efforts that aim at  unraveling the  3D  structure of
hadrons.  It should be pointed out that  the leading laboratories, like CERN (see Ref.  \cite{adolph2013hadron}, for 
recent COMPASS results) and JLAB (see, e.g., Ref. \cite{avakian2016studies}), as well as  the future 
Electron-Ion Collider, 
 { have dedicated program focused on the investigations of 
Semi-inclusive DIS processes, i.e. the main source of information on the above issue.} 
 
Moreover, on the theory side we mention the present attempts of 
getting parton distributions  as a limiting procedure applied to
imaginary-time lattice calculations, following the suggestion of Ref. \cite{XJIPRL13} 
(see 
Ref.\cite{AlexPRD15}, for some recent lattice calculations), as well as the
strong caveat contained in Ref. \cite{rossi2017note}. 
%
All that motivates a detailed presentation
of our novel method of solving BSE with spin degree of freedom in Minkowski
space, which in perspective could 
give some  reliable contribution to the  coherent efforts toward an investigation of 
the 3D tomography of hadrons.

 Our method is   based on  the so-called Nakanishi 
integral  representation (NIR) of the Bethe-Salpeter (BS)  amplitude 
(see, e.g.,  Ref. \cite{FSV1} 
for a recent  introduction to the issue, and references  quoted therein). 
{  This
 representation  
is given by a suitable integral of
the  Nakanishi weight-function (a real function) divided by a denominator 
depending upon 
 both the external four-momenta and the integration variables.
In this way, one has an  explicit expression of the analytic structure of the BS
amplitude, and proceeds to  formal elaborations. We anticipate that the 
validity of NIR for obtaining actual solutions of the ladder BSE, i.e. the one 
 we have
investigated, is achieved {\em a 
posteriori},
once an equivalent  generalized eigenvalue problem  is shown to admit solutions. 
 Within
the NIR approach} that will be illustrated in detail for the two-fermion case in
 what follows, 
several studies  have   been carried out. Among them, { one has to mention the
works devoted to the    
investigation of}: 
(i)    two-scalar   bound and 
zero-energy states, in ladder approximation with a massive 
  exchange
\cite{Kusaka,CK2006,FSV2,Tomio2016,FSV3}, as well as      
   two-fermion  ground states \cite{CK2010,dFSV1}; (ii)   
   two  scalars interacting  via a cross-ladder kernel
 \cite{CK2006b,gigante2017bound}. A major difference, that separates 
 the 
 above mentioned studies in two groups, is the technique to deal with the 
 BSE analytic structure 
  in momentum space. In Refs. \cite{FSV2,Tomio2016,FSV3,dFSV1,gigante2017bound}, it has been exploited
   the  light-front (LF) projection, which amounts to eliminate the relative 
   LF time,
between the two particles, by  integrating over the component $k^-=k^0 - k^3$
of the constituent relative momentum  (see Refs.
\cite{Sales:1999ec,SalPRC01,marinho2008light,Frederico:2010zh,FSV2} for details). 
 Such an elegant and physically motivated procedure, based on 
 the  non-explicitly covariant LF quantum-field theory 
 (see Ref.\cite{Brod_rev}),
 perfectly  combines with NIR, and, as discussed in the next Sections, it allows
 one to successfully deal with singularities (see Ref. \cite{Yan} for an early
 discussion of those singularities) that stem from the spin degrees of
 freedom acting in the problem. In general, the LF projection is able to
  exactly
 transform BSE in Minkowski space into a numerically affordable integral equation for the Nakanishi
 weight function, { without resorting to the so-called Wick rotation 
 \cite{wick1954properties}.} 
 { Specifically, the  ladder BSE   is transformed into a generalized
 eigenvalue-problem, where the Nakanishi-weight functions play the role of
 eigenvectors.}  Differently,  Refs. 
 \cite{CK2006,CK2006b,CK2010} adopted a  formal  elaboration   based on the covariant version
 of the LF quantum-field theory \cite{CK_rev}.
 This approach has not 
 allowed to formally identify
  the   singularities that plague BSE  with spin degrees of freedom, and
 therefore,  in this case, the eigenvalue equations to be solved are different from
 the ones we get. 
 In particular, in Ref.
 \cite{CK2010} the $0^+$ two-fermion case is studied and    
  a smoothing function is introduced for  
 achieving stable  eigenvalues (indeed the eigenvalues are 
 the coupling constants,
 as shown in what follows). { It should be pointed out   that in the 
 range of the
  binding energies explored in Ref. \cite{CK2010},  $B/m \in [0.01-0.5]$ ($m$ is
  the mass of the constituents), the eigenvalues }fully agree
 with the outcomes of our elaboration { for all the three interaction kernels considered} 
 (see \cite{dFSV1} and the next Sections). { As to the eigenvectors, 
 the only  case discussed in \cite{CK2010} is in an
 overall agreement with our results  (cf Sect. \ref{sect:results})}.

 { Aim of this work is to extend our}
 previous investigations of  the 
ladder BSE \cite{FSV1,FSV2,Tomio2016,FSV3,dFSV1,Pimentel:2016cpj} 
for a $0^+$ two-fermion system,  
  interacting 
through the exchange of massive 
 scalar, pseudoscalar or  vector bosons.
  Since   Ref. \cite{dFSV1} was  basically 
  devoted to validate our method through the 
  comparison of the obtained eigenvalues  and the ones found in the 
  literature, in the present paper we first provide the non trivial details of the
   formal approach, that can be adopted for future
  studies of systems with higher spins. { Then 
  we illustrate: i) for  each interaction above mentioned,}  physical motivations for the
   numerical dependence of   
 the binding energy upon the coupling
constant $g^2$ we have got; and  ii) 
 the  { peculiar outcomes} of the NIR+LF framework, represented by the eigenvectors of 
 our coupled integral system. Let us recall that the eigenvectors are 
the 
Nakanishi-weight functions,  { namely } the
fundamental ingredient for recovering both the full BS amplitude and  the
LF  amplitudes.
%
%
 { Furthermore, we  extend our analysis to a
  fermion\hyp{}antifermion pseudoscalar system  with a large binding energy,
  i.e. in a strongly relativistic regime.
 In particular,  after tuning both the fermion mass  and the
  mass of the exchanged vector-boson to
  the values suggested by lattice calculations, we show the LF amplitudes for 
  such a {\it mock pion}. They  feature the effects due to the spin degrees of
  freedom, and show  the peculiarity of an approach addressing BSE directly
  in Minkowski space.}  This preliminary study, {  once it will be enriched with
 suitable phenomenological features,  could be   relevant in 
  providing the 
  initial scale for evolving   pion  }  transverse-momentum distributions (TMD)
 \cite{BaccJHEP15}. 
 
The present paper is organized as follows. 
In section \ref{sect:twofermionbse}, we introduce 1) the basic equation for
the homogeneous two-fermion Bethe-Salpeter equation, with a kernel in ladder 
approximation, based on  three different kinds of massive exchanges, i.e.  scalar,  
pseudoscalar and  vector bosons; and 2) the general notation for the BS
amplitude of a
two-fermion $0^+$ state.
In section \ref{sect:nakareplf}, we illustrate the  Nakanishi integral 
representation for the $0^+$ bound state of two fermions 
and review the LF projection technique, that allows one   to formally 
infer an 
integral equation fulfilled by  the 
Nakanishi weight function.
In section \ref{endpointsing}, we introduce our distinctive method
for formally obtaining the kernel
 of the integral equation fulfilled by the Nakanishi weight-function,
  and separate out the
light-cone non-singular and singular contributions, by carefully analyzing 
the  end-point singularities, related to the spin degree of freedom of the problem we
are coping with.   
In section \ref{sect:nmethod}, we provide our numerical tools for solving the
integral equation for the Nakanishi weight function, that is formally equivalent to
get solution of  the BSE in Minkowski space.
In section \ref{sect:results}, we discuss  several numerical results for a
two-fermion $0^+$ state, ranging from the  
 dependence of the binding energy upon the coupling constant, peculiar for the three
 exchanges we consider to compute the LF amplitudes, building blocks of both
 LF distributions and  TMDs. In a
 forthcoming paper \cite{dFSV2} we aim to address phenomenological, but
 realistic
4D kernel to study TMDs 
\cite{BarPRep02}. 
In section \ref{sect:mockpion}, we present
an initial investigation of a fermion\hyp{}antifermion $0^-$
state,  featuring a {\em mock pion},
with input parameters  inspired by standard lattice calculations.
In section \ref{sect:conclusion}, we provide a summary 
and  concluding remarks to close our work.

\section{ General formalism for the two-fermion homogeneous BSE}
\label{sect:twofermionbse}

 In this Section, the general formalism adopted for obtaining actual
 solutions of the BSE for a bound system composed by two spin-$1/2$ constituents is 
 presented.
 Though the approach based on NIR is quite general, and it can be
 extended at least to BSE with analytic kernels, given our present knowledge, the ladder 
 approximation is suitable to
 start our novel investigation on fermionic  BSE, since it allows us to cope with 
 some fundamental
 subtleties without considering additional, but irrelevant for our most urgent
 aim, complications. We can 
 anticipate that the mentioned issues are related to the singularities onto 
 the light-cone \cite{Yan}, and the efforts for  elucidating them  are an
 unavoidable formal step in order to extend the NIR approach to higher spins (e.g. 
 vector constituents).
 Among the two-fermion bound systems, the simplest one to be 
 addressed is given by a  $0^+$ bound 
state, that after taking into account the intrinsic parity can be trivially
converted into a  $0^-$  fermion\hyp{}antifermion composite state, once  the charge conjugation
is applied. 

In what follows,  we adopt i) the  ladder approximation for the interaction
kernels, modeling  scalar, pseudoscalar or  vector exchanges, and  ii)  no  self-energy and vertex corrections,  apart a scalar form
factor to be attached at the interaction vertexes (see below). With those
assumptions,
the fermion\hyp{}antifermion BS amplitude,  $\Phi(k,p)$ 
fulfills the following integral equation \cite{CK2010}
\be
\Phi(k,p)=~S(k+p/2)~\int \frac{d^4k'}{(2\pi)^4} ~F^2(k-k') ~\nonu \times
 i{\cal K}(k,k')~\Gamma_1~\Phi(k',p)~\widehat \Gamma_2~
S(k-p/2)\, ,
\label{bse}
\end{eqnarray}
where the  off-mass-shell constituents have  four\hyp{}momenta given by
$p_{1(2)}=p/2 \pm k$, with $p^2_{1(2)}\ne m^2$, $p=p_1+p_2$ is the total 
momentum, with $M^2=p^2$   the bound-state square 
mass, and $k=(p_1-p_2)/2$ the relative four-momentum. The Dirac propagator is given by
\be
S(k)=~i~{\psla k +m\over k^2-m^2+i\epsilon}~~,
\ee
and  $\Gamma_i$ are the  Dirac structures of the
interaction vertex we will consider in what follows,  namely
 $\Gamma_i\equiv I,~\gamma_5,~\gamma^\mu$, for scalar,
pseudoscalar and  vector interactions, respectively. Moreover, 
using the charge-conjugation $4 \times 4$ matrix { $C=i\gamma^2\gamma^0$},
we define $\widehat\Gamma_2=C\,\Gamma_2^T\,C$, and 
\be
F(k-k')= {\mu^2-\Lambda^2\over (k-k')^2-\Lambda^2 +i\epsilon}
\label{vertexff}
\ee is a suitable
 interaction-vertex form factor. Besides the Dirac structure, the interaction
 vertex contains also a momentum dependence (due to the exchanged-boson
 propagation) as well as a coupling constant.
 In particular, depending on the interaction, one has 
  the following expression for $i{\cal K}$ in Eq. (\ref{bse})
\begin{itemize}
\item  for the scalar case

\be
i{\cal K}^{(Ld)}_S(k,k') = -i g^2~
{1\over (k-k')^2-\mu^2+i\epsilon},
\label{kerns}\ee
\item for the pseudoscalar one,
\be
i{\cal K}^{(Ld)}_{PS}(k,k') = i 
g^2~{1\over (k-k')^2-\mu^2+i\epsilon},
\label{kernp}\ee

\item and finally for a  vector exchange, in the Feynman gauge,

\be
i{\cal K}^{(Ld)\mu\nu}_{V}(k,k') = - ig^2~{g^{\mu\nu}\over (k-k')^2-\mu^2+i\epsilon}.
\label{kernv}\ee
\end{itemize}

The  BS amplitude $\Phi(k,p)$ 
can be decomposed as follows  \cite{CK2010}
\be
\Phi(k,p)= \sum_{i=1}^{4} S_i (k,p) ~\phi_i (k,p)~,
\label{bsa}
\ee
where 
 $\phi_i$ are suitable scalar functions of $(k^2,\, p^2,\, k\cdot p) $  
 with well-defined properties under the exchange $k
 \to -k$, namely they have to be  even for $i=1,2,4$ and odd for $i=3$. The allowed Dirac structures are given 
 by the $4\times 4$ matrices $S_i$, viz
\be
S_{1}(k,p) = \gamma_5\,,\, S_{2}(k,p) = {\psla p\over M}  \,\gamma_5\,,\,\nonu
S_{3}(k,p) = {k \cdot p \over M^3}  \psla p ~\gamma_5 - {1\over M} \psla k 
\gamma_5~, \nonu 
S_{4} (k,p)= {i \over M^2} \sigma^{\mu\nu}  p_{\mu} k_{\nu} ~\gamma_5 ~.
\label{S_structure}
\ee
They satisfy the following orthogonality relation
\be
Tr\Bigl[S_i(k,p)~S_j(k,p)\Bigr]={\cal N}_i(k,p)~\delta_{ij}.\label{normtr}
\ee
Multiplying both sides of Eq. (\ref{bse}) by $S_{i}$ and carrying out the 
traces, one reduces to a system of four coupled integral 
equations, written as
\be
\phi_i(k,p)= ig^2 (\mu^2-\Lambda^2)^2 \sum_{j}~\int {d^4k''
\over (2 \pi)^4} \nonu \times{ c_{ij}(k,k'',p)\over \Bigl[({p\over
2}+k)^2-m^2+i\epsilon\Bigr]\,\Bigl[({p\over
2}-k)^2-m^2+i\epsilon\Bigr] }
 \nonu
 \times {\phi_j(k'',p) \over \Bigl[(k-k'' )^2 -\mu^2+i\epsilon\Bigr]\,
 \Bigl[ (k-k'')^2 -\Lambda^2+i\epsilon\Bigr]^2}~,\nonu
\label{coupls}\ee
with $i,j=1,~2,~3,~4$. The coefficients $c_{ij}$  are obtained by performing the convenient traces:
\be
c_{ij}(k,k'',p)={1 \over {\cal N}_i(k,p)}~Tr\left\{S_i(k,p) 
\left(  {\psla p\over 2} +\psla k+m\right)   \right.\nonu \times
~\Gamma_1\left. ~S_j(k'',p) \widehat \Gamma_2
\left({\psla p\over 2}-\psla k-m\right )
\right\}
\label{cij1}\ee
 and are explicitly given in  \ref{coeff} (see also Ref. \cite{CK2010}). Notice
 that the coefficients for a pseudoscalar exchange and a  vector  can
 be easily obtained from the coefficients for a scalar interaction, as explained
 in   \ref{coeff}. 

\section{Nakanishi integral representation and the light-front projection}
\label{sect:nakareplf}

In the sixties, for a bosonic case, Nakanishi (see Ref. \cite{nak71} for all the details) 
proposed and elaborated  a new  integral  
representation 
for perturbative transition amplitudes,
 relying on the parametric expression of the Feynman diagrams.
  The key point in his formal approach is
 the possibility to express a $n$-leg transition amplitude as a proper folding
 of a weight function and a denominator that contains all the independent scalar
 products of the $n$ external four-momenta. Noteworthy, by using NIR, the
 analytic properties of the transition amplitudes are dictated by the above
 mentioned denominator. As a final remark, one should remind that 
 the weight function is unique, as demonstrated by Nakanishi exploiting 
  the analyticity of the 
 transition amplitude, expressed through NIR
 (see \cite{nak71}). 
 
 A step forward of topical interest was carried out in Ref. \cite{CK2010}, where
 the  generalization
 to the fermionic ground state was presented. It should be reminded that originally NIR
 was established only for the bosonic case, with some caveat about a
 straightforward application to the fermions, as recognized by Nakanishi himself, that was
 aware of the possible tricky role of the numerator in the Dirac propagator.
   
 Following the spirit of Ref. \cite{CK2010}, one can apply NIR to each scalar
 function in Eq. (\ref{bsa}), tentatively generalizing the Nakanishi approach to the fermionic case. 
 Let us recall that the denominator of a generic Feynman diagram contributing to the fermionic transition amplitudes 
 has the same expression as in the boson case analyzed by Nakanishi \cite{Nakarev}, and this is the main feature leading to NIR.
 In conclusion, one can write for each $\phi_i(k,p)$ in Eq. (\ref{bsa})
 
{\begin{small}\be
\phi_i(k,p)=
\int_{-1}^1 dz'\int_0^\infty d\gamma' 
{ g_{i}(\gamma',z';\kappa^2) \over 
\left[{k}^2+z' p\cdot k -\gamma'-\kappa^2+i\epsilon\right]^3},\nonu
\label{phinak}\ee
\end{small}}
where $\kappa^2 = m^2- M^2/4$. For each scalar  function of the BS amplitude it is associated one weight function or Nakanishi 
amplitude $g_{i}(\gamma',z';\kappa^2)$, which  is conjectured to be unique and encodes all the non perturbative dynamical information. 
The power of the denominator in Eq. (\ref{phinak}) can be chosen as any convenient integer. Actually,
the power 3 is  adopted following Ref.
 \cite{CK2010}.  The scalar
 functions $\phi_i(k,p)$ must have well-defined properties under the exchange $k
 \to -k$: even for $i=1,2,4$ and odd for $i=3$. Those properties can
  be straightforwardly
 translated to the corresponding properties of the Nakanishi weight-function 
 $g_{i}(\gamma',z';\kappa^2) $ under the
 exchange $z'\to -z'$, i.e. they must be even for $i=1,2,4$ and odd for $i=3$.

By inserting Eq. (\ref{phinak}) in Eq. (\ref{coupls}), one can write the 
fermionic BSE as a 
system of coupled integral equations, given by 

\begin{small}\be
\int_{-1}^1 dz'\int_0^\infty d\gamma' 
{ g_{i}(\gamma',z';\kappa^2)\over \left[{k}^2+z' p\cdot k
-\gamma'-\kappa^2+i\epsilon\right]^3}= \nonu =
g^2
~\sum_{j}  \int_{-1}^1 dz'  \int_0^\infty d\gamma'  
~{ K}_{ij} (k,p;\gamma', z')~g_{j}(\gamma',z';\kappa^2) ,\nonu
\label{coupls1}\ee
\end{small}
where the kernel  that includes also the Nakanishi denominator of the BS
amplitudes on the rhs is

\begin{small}
\be \label{Kij}
{ K}_{ij} (k,p;\gamma', z') = i (\mu^2-\Lambda^2)^2  
\nonu \times \int {d^4k''
\over (2 \pi)^4}~{ c_{ij}(k,k'',p)\over \Bigl[({p\over
2}+k)^2-m^2+i\epsilon\Bigr] ~ \Bigl[({p\over
2}-k)^2-m^2+i\epsilon\Bigr]}
 \nonu ~\times
{1\over\Bigl[(k-k'')^2 -\Lambda^2+i\epsilon\Bigr]^2 
~\Bigl[(k-k'' )^2 -\mu^2+i\epsilon \Bigr]}
\nonu \times
{1\over  \Bigl[{k''}^2+z' p\cdot k'' -\gamma'-\kappa^2+i\epsilon\Bigr]^3    
} ~.
\ee
\end{small}
It is necessary to stress that the validity of the NIR for the
BS amplitude is verified {\em a posteriori}. Namely, if the generalized
eigen-equation in Eq. (\ref{coupls1}) admits eigen-solutions then NIR can be certainly applied to
the scalar function $\phi_i(k,p)$. Let us recall that Eq. (\ref{coupls}) formally 
follows  from
the BSE in Eq. (\ref{bse}).

One can perform the
four-dimensional integration on $k''$ in Eq. (\ref{Kij}), obtaining

\begin{small}
\be
{ K}_{ij} (k,p;\gamma', z') =  
\nonu = {1\over 8\pi^2 M^2}~
~ (\mu^2-\Lambda^2)^2
~\sum_{n=1}^{3} {\cal P}^{(n)}_{ij}(k,p;\gamma', z')  ~, \nonu
\label{calk2}
\ee
\end{small}
where

\begin{small}
\be
{\cal P}^{(1)}_{ij}(k,p;\gamma', z')= \nonu
= {a^0_{ij}+ a^1_{ij}~(p\cdot k)+a^2_{ij}~(p\cdot k)^2+a^3_{ij}~k^2\over
[(1-z)~(k^- - k^-_d) +i\epsilon]~[(1+z)~(k^- - k^-_u) -i\epsilon]}
\nonu \times ~
\int^1_0 dv~v^2~(1-v)^2~F(k^-,\gamma,z;\gamma', z';v)
\label{calp1}\ee

\be
{\cal P}^{(2)}_{ij}(k,p;\gamma', z')= \nonu
= {b^0_{ij}+ b^1_{ij}~(p\cdot k)+b^2_{ij}~(p\cdot k)^2+b^3_{ij}~k^2 \over 
~
[(1-z)~(k^- - k^-_d) +i\epsilon]~[(1+z)~(k^- - k^-_u) -i\epsilon]} 
\nonu \times ~
\int^1_0 dv~v^2~(1-v)^3 ~F(k^-,\gamma,z;\gamma', z';v)
\label{calp2}\ee
\be
{\cal P}^{(3)}_{ij}(k,p;\gamma', z')=\nonu
 ={d^0_{ij}+ d^1_{ij}~(p\cdot k)
\over
[(1-z)~(k^- - k^-_d) +i\epsilon]~[(1+z)~(k^- - k^-_u) -i\epsilon]} \nonu
\times 
\int^1_0 dv~v^2~(1-v)^3
\Bigl[(p\cdot k)^2-M^2 k^2
\Bigr] \nonu \times~F(k^-,\gamma,z;\gamma', z';v) ~~,
\label{calp3} 
\ee
\end{small}
and the coefficients $a^k_{ij}$, $b^k_{ij}$ and $d^k_{ij}$ are given in  \ref{coeff}. In the above equations, one has
\be
k^-_d=-{M\over 2}+ {2 \over M(1-z)} (\gamma+m^2)
\nonu
k^-_u={M\over 2}- {2 \over M(1+z)} (\gamma+m^2)
\ee
where the the LF variables $k^\pm=k^0\pm k^3$  
have
been used, and the following notation has been adopted: $\gamma=|{\bf k}_\perp|^2$ and  $z=-2k^+/M$ $\in[-1,1]$ (see Ref.
\cite{FSV1} for details on this range).
Moreover,
 in Eqs. (\ref{calp1}), (\ref{calp2}) and (\ref{calp3}), one has
 
\begin{small}
 \be
 F(k^-,\gamma,z;\gamma', z';v)= {1\over \Bigl[k^-~k^+_D +\ell_D
+i\epsilon\Bigr]^2}
   \nonu
 {3
 (k^-~k^+_D +\ell_D)
 +(1-v)\Bigl ( \mu^2 -\Lambda^2\Bigr) \over \Bigl[k^-~k^+_D +\ell_D
+(1-v)\Bigl ( \mu^2 -\Lambda^2\Bigr)+i\epsilon\Bigr]^3}
 \label{capf}\ee
\end{small}
with 

\begin{small}
\be
k^+_D=v(1-v){M\over 2}~(z'-z)
\nonu
\ell_D=
-v(1-v) \Bigl(\gamma+zz'{M^2\over 4}-{z'}^2 {M^2\over 4}\Bigr)
\nonu
-v\Bigl[\gamma'  +z'^2m^2+ (1-z'^2)\kappa^2\Bigr] -(1-v)\mu^2~~.
\label{ellD}\ee
\end{small}

By a direct inspection of Eq. (\ref{capf}), one can realize that several poles affect 
the needed integration on the relative four-momentum.
 In order to treat the poles in a proper way,  we adopt the so-called 
 LF-projection onto the null-plane 
 (see, e.g.\cite{Sales:1999ec,Frederico:2010zh,marinho2008light}), that amounts to 
 integrate over the LF component $k^-$ both sides of  Eq. (\ref{coupls1}). It should be pointed out that such a
 LF-projection is  specific of the non-explicitly covariant version of the LF
 framework, and it is a key ingredient in our approach, as it will be clear in
 the next Sections. As to the present stage of the elaboration, 
 the LF projection
    marks
 one of the differences with
  the explicitly covariant approach in Ref.
 \cite{CK2010}. 
 
 To conclude this Section, it should be emphasized that
 the announced difficulties in the fermionic case are related to
the possible light-cone singularities generated by powers of $k^-$
contained  in 
the coefficients $c_{ij}(k,k'',p)$ (see also Ref. \cite{Yan}
  and Refs. \cite{Bakker:2007ota,bakker2005restoring,Melikhov:2002mp}
  for more details on the light-cone singularities).
In the next Sections, the singularities will be identified and formally 
integrated out.

\subsection{Light-front projection of the fermionic Bethe-Salpeter equation}

The integration of the lhs is straightforward, since it is analogous to the scalar case.  After the LF projection one gets {what we shortly call 
 {\em light-front amplitudes}
\be
\label{psii}
\psi_{i} (\gamma,\xi)=\int {dk^-\over 2\pi}~\phi_i(k,p)=\nonu
= -{i \over M} \int_0^\infty d\gamma' 
{ g_{i}(\gamma',z;\kappa^2) \over \left[\gamma + \gamma' + m^2 z^2 + 
(1-z^2)\kappa^2 -i\epsilon
\right]^2
}\nonu
\end{eqnarray}
where  $\xi=(1-z)/2$  belongs to  $[0,1]$. Notice 
that  $\psi_i$ are  scalar 
functions rotationally invariant into the transverse plane  and they
will be used for constructing \cite{dFSV2} the so-called valence component of the two-fermion
state, once a Fock expansion of this state is introduced. In Eq. (\ref{psii}), 
$i\epsilon$ can be removed,
since we are dealing with a bound state and
 $\kappa^2>0$.

Differently, the integration on $k^-$ of the rhs  of  Eq. (\ref{coupls1}) 
has to be done very carefully and it represents
the core of our approach. A first, but short, presentation can be found in Ref.
\cite{dFSV1}, together with
some important numerical outcomes.

Then, performing the LF projection as in Eq. (\ref{psii}), one can  
 transform Eq. (\ref{coupls1}) into a new coupled integral\hyp{}equation system for the
 Nakanishi weight\hyp{}functions, viz
 
\begin{small}\be
\label{coupls2}
\int_0^\infty d\gamma' { g_{i}(\gamma',z;\kappa^2) \over 
\left[\gamma + \gamma' + m^2 z^2 + (1-z^2)\kappa^2\right]^2} = 
iM g^2\nonu
\times ~\sum_{j}  \int_0^\infty d\gamma' \int_{-1}^1 dz'{\cal L}_{ij} 
(\gamma,z;\gamma', z') 
 ~g_{j}(\gamma',z';\kappa^2)\, , \nonu
\ee
\end{small}
where

\begin{small}
\be
{\cal L}_{ij} (\gamma,z;\gamma', z') =  \nonu=
{ (\mu^2-\Lambda^2)^2\over 8\pi^2 M^2}
 ~\int {dk^- \over 2 \pi} \sum_{n=1}^{3} {\cal P}^{(n)}_{ij}(k,p;\gamma', z') 
 ~,
\label{Lij}
\ee 
\end{small}
This coupled system has the attractive feature that one has
 to deal with a single non compact
variable, $\gamma'$, and a compact one, $z'$ (like for  $\gamma$ and $z$,
respectively).

In the next subsection the integration on $k^-$ in Eq. (\ref{Lij}) will be 
discussed in detail.

\subsection{The kernel operator of the coupled integral-equation system}
One can write the kernel ${\cal L}_{ij}$ as follows
\be 
{\mathcal L}_{ij} (\gamma,z;\gamma', z') =  {(\mu^2-\Lambda^2)^2\over 8 \pi^2 M^2} 
\int^1_0 dv~v^2~(1-v)^2 \, \nonu \times
\sum_{n=0}^{3} F_{n;ij}(v,\gamma,z,p)~{\cal C}_{n} \label{Lij1}
\ee
where the non vanishing $F_{0;ij} $, $F_{1;ij} $, $F_{2;ij} $ and $F_{3;ij} $ 
are listed in Table \ref{F0}.

\begin{table*}
\begin{center}
 \caption{Non vanishing coefficients $F_{0;ij}$, $F_{1;ij}$, $F_{2;ij}$ and $F_{3;ij}$.
 } \label{F0}
 \begin{tabular}{|c|c|c|c|c|}
 \hline
 $~ij~$&$F_{0;ij}$&$F_{1;ij}$&$F_{2;ij}$&$F_{3;ij}$\\
 \hline
$11$ & $ m^2+M^2/4+\gamma $ & $  z M/2  $ & $ 0  $ & $  0 $\\
$12$ & $ mM $                            & $  0  $ & $ 0  $ & $  0 $\\
$14$ & $ - (1-v)\left( \gamma +z^2{M^2\over 16}\right) $ & $ -(1-v) zM/4   $ & $ -(1-v)/4   $ & $  0 $\\
$21$ & $ mM $ & $  0  $ & $ 0  $ & $  0 $\\
$22$ & $ m^2+M^2/4-\gamma-z^2{M^2/ 8}  $ & $  0  $ & $ -(1/2)   $ & $  0 $\\
$23$ & $ (1-v)z \left(\gamma  +z^2{M^2\over 16}\right) /2 $ & $  -(1-v)\left(\gamma  -z^2{M^2\over 16}\right) /M  $ & $ -(1-v)z/8   $ & $  -(1-v)/(4M)  $\\
$24$ & $ - (1-v)\left( \gamma +z^2{M^2\over 16}\right)2m/(M)$ & $  -(1-v) zm/2  $ & $ -(1-v)m/(2M)   $ & $  0 $\\
$32$ & $ -z{M^2/ 2}$ & $  M  $ & $ 0  $ & $  0 $\\
$33$ & $ (1-v)\Bigl[m^2-M^2/4+\gamma+z^2{M^2/ 8}\Bigr]$ & $  0  $ & $ (1-v)/2  $ & $  0 $\\
$34$ & $ - (1-v)z mM/ 2$ & $  (1-v) m  $ & $ 0  $ & $  0 $\\
$41$ & $M^2$ & $  0  $ & $ 0  $ & $  0 $\\
$42$ & $ 2mM$ & $  0  $ & $ 0  $ & $  0 $\\
$43$ & $ - (1-v)z mM/ 2 $ & $  (1-v) m  $ & $ 0  $ & $  0 $\\
$44$ & $  (1-v)\Bigl[m^2-M^2/4-\gamma\Bigr] $ & $  -(1-v) zM/2  $ & $ 0  $ & $  0 $\\
 \hline
 \end{tabular}
 \end{center}
 \end{table*} 

Moreover, in Eq. (\ref{Lij1}) the functions 
${\cal C}_n$ are integrals over $k^-$ defined by

\begin{small}
\be
{\cal C}_n =
\int {dk^-\over 2 \pi}
 {(k^-)^n ~F(k^-,\gamma,z;\gamma',z';v)
\over
 \Bigl[(1-z)k^- -(1-z) k^-_d+i\epsilon\Bigr]}\nonu \times~{1
\over
\Bigl[(1+z)k^- -(1+z)k^-_u-i\epsilon\Bigr]
}
 =
\nonu=
-{\partial \over \partial \ell_D}\int {dk^-\over 2 \pi}
{ (k^-)^n \over
 \Bigl[(1-z)k^- -(1-z) k^-_d+i\epsilon\Bigr]} \nonu \times~
{1\over \Bigl[(1+z)k^- -(1+z)k^-_u-i\epsilon\Bigr]
}
\nonu
 \times
 ~{1\over
 \Bigl[k^+_D k^-+\ell_D
+(1-v)\Bigl ( \mu^2 -\Lambda^2\Bigr)+i\epsilon\Bigr]^2} \nonu \times ~
{1\over \Bigl[k^+_D k^-+\ell_D
+i\epsilon\Bigr]}~,
\label{cn}
\ee
\end{small}
with  $n=0,1,2,3$.

It is fundamental to notice that the powers of $k^-$ in the numerator of Eq.
(\ref{cn}) are dictated by the coefficients $c_{ij}(k,k'',p)$ 
(see Eq. (\ref{cij1})), that in turn are
determined by the Dirac structures of both the BS amplitude (cf Eq. (\ref{bsa})) 
and the interaction vertexes  (cf, e.g., the matrix  $\Gamma_i$ in the ladder
 BSE in Eq. (\ref{bse})). Since  the
power of $k^-$ in the numerator lead to singular integrals, as
discussed below, one has to  bring in mind that the most severe singularities
can be determined in advance, once the Dirac structures above mentioned are
known. Therefore, if one has to deal with the BS amplitude of a
 fermion-boson system or a
vector-vector one, the biggest power of $k^-$ can be easily singled out after
modeling 
(in agreement to the required symmetries and the interaction) the needed Dirac 
structures.

As it is clearly shown in Eq. (\ref{cn}), the $k^-$ integration of the
kernel ${\cal L}_{ij}$ could become tricky if the  powers of $k ^-$ in the
numerator are not suitably balanced by the ones in the denominator for some values of
$k^+_D$. The values $z=\pm 1$, that in principle could generate troubles, are made 
harmless by the vanishing values of the Nakanishi weight functions at those
values (see the discussion of the boundaries for  
the scattering case in Ref. \cite{FSV3}, that can be adapted to the present
discussion).
In conclusion, to perform the  projection of the kernel ${ K}_{ij}$ 
onto the LF hyper-plane, one has to carefully study the integrals ${\cal C}_n$. 
In particular,  we should analyze 
the behavior of the integrand when the contour arc goes to infinity 
in order to use the Cauchy's residue  theorem in Eq. (\ref{cn}). 
The critical situation occurs for $k^+_D = 0$, that was not recognized  
as the source
of the instabilities met in Ref. \cite{CK2010} (they were fixed only through a
numerical procedure). Indeed, to single out the actual sources of
instabilities makes more sound the NIR approach for systems with spin degrees of
freedom. 
\section{The integration on $k^{-}$ of the kernel ${\cal L}_{ij}$ and the
 LF singularities} 
\label{endpointsing}
In this Section, 
we present the detailed results for the
integration on $k^{-}$ in Eq. (\ref{cn}), considering in the next subsections  
   two different cases: i)
$k^{+}_{D}\ne 0$ that will lead us to obtain a result corresponding to  Eq.
 (15) of Ref. \cite{CK2010};
  and  ii) any $k^{+}_{D}$, i.e. including both the
 previous case and the tricky $k^+_D=0$. The analysis of the  last case  
  represents 
 our main contribution
 to the topic, since we are able both to achieve a formal result, that explains why
 in Ref. \cite{CK2010} instabilities were found, and to obtain numerical 
 results for the eigenvalues
 in full agreement with Ref. \cite{CK2010}, where the issue was fixed through
 the introduction of a smoothing function. Also the eigenvalues obtained in Ref. \cite{Dork}
 by using a Euclidean BSE are recovered within our approach.

\subsection{Non-singular kernel: $k^+_D\ne 0$}\label{nosing}

Let us first consider the case $k^+_D\ne 0$. Then, in the denominator of Eq. (\ref{cn}), for  $n=0,1,2,3$, 
one has  powers of $k^{-}$ large 
enough  so that  
the arc contribution  at infinity vanishes and one can safely apply the
Cauchy's residue theorem.
For $k^+_D>0$ one can close the path in the upper semi-plane and 
pick up the contribution of the
residue at the pole $k^-_u+i\epsilon$. It is understood that for $z \to~-1$ and
$k^+_D>0$ one has a 
vanishing result, since the remaining  poles belong  to the
same semi-plane. In this case, one gets 

\begin{small}
\be
{\cal C}^{(+)}_n(\eta)= -
i~\theta(k^+_D-\eta)~{ M\over 4}~{(k^-_{u})^n
\over
 \Bigl[\gamma +z^2m^2+(1-z^2)\kappa^2\Bigr]}
 \nonu \times
 ~F(k^-_u,\gamma,z;\gamma',z';v)~~,
\label{cn+}\ee
\end{small}
where the small, positive quantity $\eta$ has been put in the theta function in
order to strictly exclude the case $k^+_D=0$, and $\ell_D$ is given in Eq. (\ref{ellD}). Moreover, it is noteworthy that 
the denominator
$\gamma+z^2m^2+(1-z^2)\kappa^2
$ is always non vanishing for a bound state, since $\kappa^2\ge 0$.

Analogously, for $k^+_D<0$ we can close the path in the lower semi-plane and
 pick up the residue contribution at
$k^-_d-i\epsilon$,  obtaining

\begin{small}
 \be
{\cal C}^{(-)}_n(\eta)
=-i~\theta(-k^+_D-\eta)~
{ M\over 4}~{(k^-_{d})^n
\over
 \Bigl[\gamma+z^2m^2+(1-z^2)\kappa^2  \Bigr]
}
 \nonu
 \times
  ~F(k^-_d,\gamma,z;\gamma',z';v)~~.
\ee
\end{small}
If the condition $k^{+}_{D} \neq 0$ is satisfied, by defining 
\be
{\cal C}^{(ns)}_n=\lim_{\eta\to 0} \Bigl[{\cal C}^{(+)}_n(\eta) + {\cal C}^{(-)}_n(\eta)
 \Bigr] 
\label{cnns}\ee
one can get
what  we call non singular contribution to the kernel ${\cal L}_{ij}$  i.e.

\begin{small}
\be \label{LijNS}
{\cal L}^{(ns)}_{ij} (\gamma,z;\gamma', z') = {(\mu^2-\Lambda^2)^2\over 8 \pi^2 M^2} \times
~\nonu
\int^1_0 dv~v^2~(1-v)^2 \, \sum_{n=0}^{3} F_{n;ij}(v,\gamma,z,p)~{\cal C}^{(ns)}_n~.
\ee
\end{small}
More explicitly

\begin{small}
\be
{\cal L}^{(ns)}_{ij}(\gamma,z;\gamma',z')= 
~-{i\over M} {(\mu^2-\Lambda^2)^2\over 32 \pi^2 }\nonu \times~
{1\over
\Bigl[\gamma+z^2m^2+(1-z^2)\kappa^2  \Bigr]}~
\int^1_0 dv~v^2~(1-v)^2\nonu \times ~
\Bigl\{\theta(k^+_D-\eta)
 ~ {\cal F}_{ij} (v,\gamma,z,k^-_{u},p)
 \nonu \times F(k^-u,\gamma,z;\gamma',z,;v)
  + \sigma_{ij}~\Bigl[ z\to -z; ~z' \to -z'\Bigr]\Bigr\},
\nonu\label{nonsing}\ee
\end{small}
where  $\sigma_{ij}$ reads (cf Ref. \cite{CK2010})
\begin{center}
$\sigma = \left(
\begin{array}{cccc}
 +1 & +1 & -1 & +1 \\
 +1 & +1 & -1 & +1 \\
 -1 & -1 & +1 & -1 \\
 +1 & +1 & -1 & +1 \\
\end{array}
\right)~~~~~,$
\end{center}
and
\be
{\cal F}_{ij} (v,\gamma,z,k^-_{u},p)= \sum_{n=0}^{3} (k^-_{u})^{n} F_{n;ij} (v,\gamma,z) ~.
\ee

 The coefficients  $C_{ij}(\gamma,z;v)$ 
in Ref. \cite{CK2010} are related to 
${\cal F}_{ij} (v,\gamma,z,k^-_{d},p)$   as
follows
\be
C_{ij}(\gamma,z;v)={ {\cal F}_{ij} (v,\gamma,z,k^-_{d},p) \over 4m^2} ~.
\ee
The kernel ${\cal L}^{(ns)}_{ij}$, inserted in Eq. (\ref{coupls2}),  exactly leads
  to 
 Eq. (15) of Ref. \cite{CK2010}, though  Eq. (\ref{nonsing}) has 
 been  obtained within the non-explicitly covariant
version of the LF approach, while  Ref. \cite{CK2010} has  exploited the
explicitly covariant one (see, e.g., Ref. \cite{CK_rev}).

\subsection{The $k^-$ integration of the kernel 
${\cal L}_{ij}$ for any $k^+_D$}\label{anykD}
For vanishing $k^+_D$ the powers of $k^-$ in the numerator of ${\cal L}_{ij}$ make the analytic
 integration tricky, since one cannot naively close the
 integration path with an arc at infinity.
In  \ref{genint}, all the needed details are given, while in this
Subsection the results  are summarized and commented.

The main formal tool we have exploited is given by the following singular
integral studied in detail by Yan  in Ref. \cite{Yan},
that, indeed, belongs to a series of papers devoted to  the  analysis of the field
 theory in the infinite momentum frame. The mentioned integral is
 \be \label{yan}
\int_{-\infty}^\infty {dx\over 2\pi} ~
{ 1 \over \Bigl[\beta~x -y\mp i \epsilon\Bigr]^2}= ~\pm i
~{\delta (\beta)\over \Bigl[-y\mp i \epsilon\Bigr]}~~.
\ee

 In order to complete our analysis we also exploited
  the derivative with respect to
 $y$, i.e.
 
 \begin{small}
 \be
 \int_{-\infty}^\infty {dx\over 2\pi} ~
{ 1 \over \Bigl[\beta~x -y\mp i \epsilon\Bigr]^n}= \pm  ~{i\over n-1}
~{\delta (\beta)\over \Bigl[-y\mp i \epsilon\Bigr]^{n-1}}  ~,
\nonu\ee
\end{small}
with $n>2$. Indeed, in what follows, only first ($n=3$) and second ($n=4$)
derivatives are used.

Then, the function ${\cal C}_n$ can be decomposed as follows
\be
{\cal C}_0={\cal C}^{(ns)}_0
\nonu
{\cal C}_1={\cal C}^{(ns)}_1
\nonu
{\cal C}_2={\cal C}^{(ns)}_2+{\cal C}^{(s)}_2
\nonu
{\cal C}_3={\cal C}^{(ns)}_3+{\cal C}^{(s)}_3~~,
\label{cngena}\ee
with ${\cal C}^{(ns)}_n$ as given in Eq. (\ref {cnns}) and (see  \ref{genint})

\begin{small}\be
{\cal C}^{(s)}_2=-{i \over (1-z^2)}
~{\delta(k^+_D)\over \ell_D
\Bigl[\ell_D 
+ (1-v)\Bigl ( \mu^2 -\Lambda^2\Bigr) +
i\epsilon\Bigr]^2}~,
\nonu
{\cal C}^{(s)}_3={i \over (1-z^2)}~{\partial \over \partial k^+_D}
\delta(k^+_D)~{1 \over \Bigl[(1-v)\Bigl ( \mu^2 -\Lambda^2\Bigr)\Bigr]^2}
\nonu \times~\left[ \ln\left({\ell_D +  (1-v) (\mu^2-\Lambda^2)\over \ell_D}\right)
\right. \nonu \left.-{ (1-v) (\mu^2-\Lambda^2)
\over \ell_D +  (1-v) (\mu^2-\Lambda^2)}\right]
\nonu
-{i \over (1-z^2)}~{(k^-_u+k^-_d)
~\delta(k^+_D)~\over \ell_D
\Bigl[\ell_D 
+ (1-v)\Bigl ( \mu^2 -\Lambda^2\Bigr) +
i\epsilon\Bigr]^2}~~.\nonu
\ee
\end{small}
According to the above decomposition,  the kernel 
${\cal L}_{ij}$ in Eq. (\ref{Lij}) can be written
\be
{\cal L}_{ij}(\gamma,z;\gamma',z')= {\cal L}^{(ns)}_{ij}(\gamma,z;\gamma',z')
+{\cal L}^{(s)}_{ij}(\gamma,z;\gamma',z')\nonu
\label{decLij}
\ee
where ${\cal L}^{(ns)}_{ij}$ is given in Eq. (\ref{LijNS}) and 
${\cal L}^{(s)}_{ij}$ is 
\be
{\cal L}^{(s)}_{ij}(\gamma,z;\gamma',z')=
{(\mu^2-\Lambda^2)^2\over 8 \pi^2M^2}
\int^1_0 dv~v^2~(1-v)^2\nonu \times ~\Bigl[ F_{2;ij}(v,\gamma,z)
~{\cal C}^{(s)}_2+F_{3;ij}(v,\gamma,z)~{\cal C}^{(s)}_3\Bigr]~~.
 \label{LijS}
\ee
Due to  the factor $v^2(1-v)^2$ in Eq. (\ref{LijS}), that helps us to get rid of
the end-point issues at $v=0$ and $v=1$ one can safely write
\be
\delta(k_D^+) = \frac{2~\delta(z'-z)}{M~ v(1-v)}
\nonu
{\partial \over \partial k^+_D}\delta(k_D^+) = \frac{4}{[ M~v(1-v)]^2}~
\Bigl[{\partial \over \partial z'}\delta(z'-z)\Bigr]
\ee
obtaining
\be
{\cal L}^{(s)}_{ij}(\gamma,z;\gamma',z')=
-i~{ 2\over M(1-z^2)}{(\mu^2-\Lambda^2)^2\over 8 \pi^2M^2}
\int^1_0 dv\nonu\times ~v(1-v) 
 ~\Bigl\{{\delta(z-z')~\over \ell_D
\Bigl[\ell_D 
+ (1-v)\Bigl ( \mu^2 -\Lambda^2\Bigr) +
i\epsilon\Bigr]^2}
\nonu \times\Bigl[F_{2;ij}(v,\gamma,z)+(k^-_u+k^-_d)
F_{3;ij}(v,\gamma,z)\Bigr] 
\nonu
-F_{3;ij}(v,\gamma,z)~\frac{2}{ M~v(1-v)}~
\Bigl[{\partial \over \partial z'}\delta(z'-z)\Bigr] ~{\cal D}^{(s)}_3
\Bigr\}~~,
\nonu \label{LijS1}
\ee
with 
\be
{\cal D}^{(s)}_3=
{1 \over \Bigl[(1-v)  \Bigl ( \mu^2 -\Lambda^2\Bigr)\Bigr]^2} \nonu \times
\left[ \ln\left({\ell_D +  (1-v)(\mu^2-\Lambda^2)\over \ell_D}\right)
\right. \nonu \left. -{ (1-v) (\mu^2-\Lambda^2)
\over \ell_D + (1-v) (\mu^2-\Lambda^2)}\right]~.
\ee

The following physical interpretation should be associated with the presence of singular contributions to the kernel
of the integral equation when $k^+_D$ vanishes, since it is proportional to the  plus momentum carried out by the exchanged quantum. 
Note that $k^+_D\propto (z^\prime-z)$, where the momentum fractions of the external  and internal fermions are $0<\xi =(1-z)/2<1$ and 
 $0<\xi^\prime =(1-z^\prime)/2<1$, respectively. 
Hence, one can identify the contribution of a LF zero-mode to the kernel when the
exchanged quanta propagates along the light-like direction ($k^+_D=0$). 
In some cases it can give rise to a LF singularity, as mentioned above. 

Of course, the problem is avoided if the fermions propagate before the exchange of the 
light-like quanta. The high virtuality of the quanta with $k^+_D=0$ in general suppresses 
any intermediate propagation, as for example 
in the two-scalar BSE, 
unless the exchanged boson couples with higher spin particles. This happens when the so-called instantaneous
terms and/or derivative coupling are considered.

In conclusion, the final form of the homogeneous ladder BSE for two 
fermions reduces to
\be
\int_0^\infty d\gamma' 
{ g_{i}(\gamma',z;\kappa^2) \over \left[\gamma + \gamma' + m^2 z^2 + (1-z^2)\kappa^2\right]^2}
=
\nonu= iM g^2 \sum_{j}~\int_{-1}^1 dz'\int_0^\infty d\gamma' ~ 
\Bigl[{\cal L}^{(ns)}_{ij}(\gamma, z, \gamma', z') \nonu
 + {\cal L}^{(s)}_{ij}(\gamma, z, \gamma', z')\Bigr]~g_{j}(\gamma',z';\kappa^2)~, 
\label{bsefin}
\ee
where ${\cal L}^{(ns)}_{ij}(\gamma, z, \gamma', z')$ and ${\cal L}^{(s)}_{ij} $
 are given by Eqs. (\ref{LijNS}) and (\ref{LijS}), respectively.

In physical terms the essential contribution of the instantaneous terms 
and the light-like exchanged quanta 
should be associated with a two-fermion probability amplitude equally distributed along the light-like trajectory 
contained in the null-plane.
 
\section{Numerical Method}\label{sect:nmethod} 
For the $0^+$ two-fermion bound state, the system in Eq. (\ref{bsefin})
contains four coupled integral equations. 
In order to solve such a system, we introduce 
 a basis expansion for the Nakanishi weight function
\be
g_{i}(\gamma, z) = \sum_{m=0}^{\infty} \sum_{n=0}^{\infty}
w_{mn}^{i}G^{\lambda_i}_{2m+r_i}(z) \, \mathcal{J}_{n}(\gamma) ~,
\label{exp}\ee
where i) $w_{mn}^{i}$ are suitable coefficients to be determined by solving the
generalized eigenproblem given by the coupled-system of integral equations  (\ref{bsefin}), ii) 
$G^{\lambda_i}_{2m+r_i}(z)$, with $r_i=0$ for $i=1,2,4$ and $r_3=1$ (due to the
symmetry under the exchange $z\to -z$) are related to the Gegenbauer polynomials 
$C_{2m+r_i}^{\lambda_i}(z)$ and iii) $\mathcal{J}_{n}(\gamma)$  to the
 Laguerre $L_{n}(a\gamma)$. In particular, one has
\be
 G_{n}^{\lambda} (z)  =  (1-z^2)^q\, \Gamma(\lambda) 
 \sqrt{\frac{n! (n + \lambda)}{ 2^{1-2\lambda} \, \pi\, \Gamma(n + 2\lambda)}}
~ C_{n}^{\lambda}(z) ~,\nonu
\mathcal{J}_{n}(\gamma)  =  \sqrt{a} \, L_{n}(a\gamma) e^{-a\gamma/2} ~.
\ee
with $ q=(2\lambda-1)/4$.
The following orthonormality conditions are fulfilled
\be
\int^1_{-1}dz~G^{\lambda_i}_\ell(z)~G^{\lambda_i}_n(z)=\delta_{\ell n} ~,
\nonu
  \int_0^\infty~ d\gamma~{\cal J}_j(\gamma)~{\cal J}_\ell(\gamma)=
~\delta_{j\ell} ~.
\ee
Notice that the kernel ${\cal L}^{(s)}$ in Eq. (\ref{bsefin}) contains
 terms proportional to derivatives of the delta-function $ \delta' (z'-z)$  and therefore
 the numerical evaluation needs derivatives of the basis functions $G^{\lambda_i}_\ell(z)$.
 
 { After inserting the above expansion in Eq. (\ref{bsefin}), one obtains a
 matrix form of the  system, reducing to solve a {\em discrete generalized eigenvalue
 problem} as in the simple case of two scalars (see ref. \cite{FSV2}). At
 the end, the generalized eigenvalue problem to be solved looks like { $$C
 ~\vec{w}=g^2~ D~\vec{w}~~,$$ with i) $C$ and $D$ square matrices,  
 and ii) $\vec{w}$ the eigenvector that
 allows one to reconstruct the Nakanishi-weight functions (cf Eq.
 \eqref{exp}).}
 The eigenvalue is $g^2$, as in the standard way for investigating the ladder BSE,
 since the binding energy, defined by $B=2m-M$ is a non linear parameter, and it
 is assigned in the range $2\ge B/m\ge 0$, given the well-known instability that
 occurs for odd numbers of boson fields (see, e.g., \cite{Baym} for the $\phi^3$
 case).}
 
  { In Ref. \cite{dFSV1}, where the eigenvalues were shown for several
  cases,} the calculations have been carried out taking as the biggest basis the one 
with $44$ Laguerre polynomials and $44$ Gegenbauer ones with indices for each
 Nakanishi weight functions $5/2,
7/2,7/2,7/2$, respectively.  To improve  the convergence, in those calculations the parameter 
$a=6.0$ 
has been adopted, and   the variable $\gamma$ has been rescaled according to 
$\gamma 
\to 2 \gamma/a_0$  with
$a_0=12$.  The two parameters $a$ and $a_0$ help to take into account 
  the range of
relevance of the Laguerre polynomials and the structure of the kernel, 
respectively. Finally, a small quantity, $\epsilon=10^{-7}$, has been added to
the diagonal elements of the  discretized matrix on the lhs of the system in
Eq. (\ref{bsefin}).

The study of the eigenvectors is  more delicate, even if the final
goal is the calculation of the LF amplitudes, Eq. (\ref{psii}), that have a
behavior extraordinary stable against the increasing of the dimension of the
basis. Indeed, the integration on the variable $\gamma'$ acts as a filter
with respect the oscillations that could affect the Nakanishi weight functions.
To have a better numerical control on the singular contributions at the
end-points  $z=\pm 1$, we 
have multiplied both  sides of Eq. (\ref{bsefin}) by a factor $(1-z^2)^3$,
rather than increasing the index $\lambda$ of the Gegenbauer polynomials. This
second option entails an increasing of the  oscillation amplitudes of the basis in $z$.
The power of the factor $(1-z^2)$  is suggested by the quantity ${\cal L}_{23}^{(s)}(a)$ defined in Eq. (\ref{l23sab}) and entering in the expression of
${\cal L}^{(s)}_{23}(\gamma, z, \gamma', z')$ (see Eq. (\ref{decLij})).
{Finally, for the cases with $B/m>1.0$, with the same aim of increasing
the stability, we have chosen $a=3$, and only for $g_3(\gamma,z;\kappa^2)$, odd in
$z$, we have reduced the dimension of the basis from 44 to 22 Gegenbauer polynomials.} 

\section{Results for Scalar, Pseudoscalar and Vector exchanges}\label{sect:results}

 The solution of the coupled set of integral equations for the Nakanishi 
weight-functions (\ref{bsefin}) 
was obtained for three different ladder kernels 
(see Sect. \ref{sect:twofermionbse}) featuring 
scalar, pseudoscalar and  vector exchanges, besides an  interaction vertex
smoothed through a 
 form factor (Eq. (\ref{vertexff})). Our study aims at singling out 
signatures of the dynamics generated by the different kind of exchanges. Indeed,
among the physical motivations of such a broad analysis,
 we have to put the application to the study of pseudoscalar mesons, originated by  vector exchanges
 between 
 quarks, and in what follows a first attempt, that we call {\it mock pion}, 
 will be discussed (see Sect. \ref{sect:mockpion}). 
 
 { The differences among the coefficients present in the kernel of the coupled equations 
are due to the peculiar Dirac structures  entailed by  scalar,
 pseudoscalar and
vector exchanges (see Eqs. (\ref{cijpseudo}) and 
 (\ref{cijvector})). Such differences are naturally reflected 
  in the corresponding 
Nakanishi functions, since they  are  suitably weighted
in the integral equations through the above mentioned coefficients. Eventually, the
differences  show up in the LF amplitudes, Eq. (\ref{psii}).}

In this section we focus on the scalar,  pseudoscalar and vector cases, leaving
the actual application to the {\em mock pion} to the next Section.

\begin{figure}[thb] 
\centering
   \includegraphics[width=9.0cm]{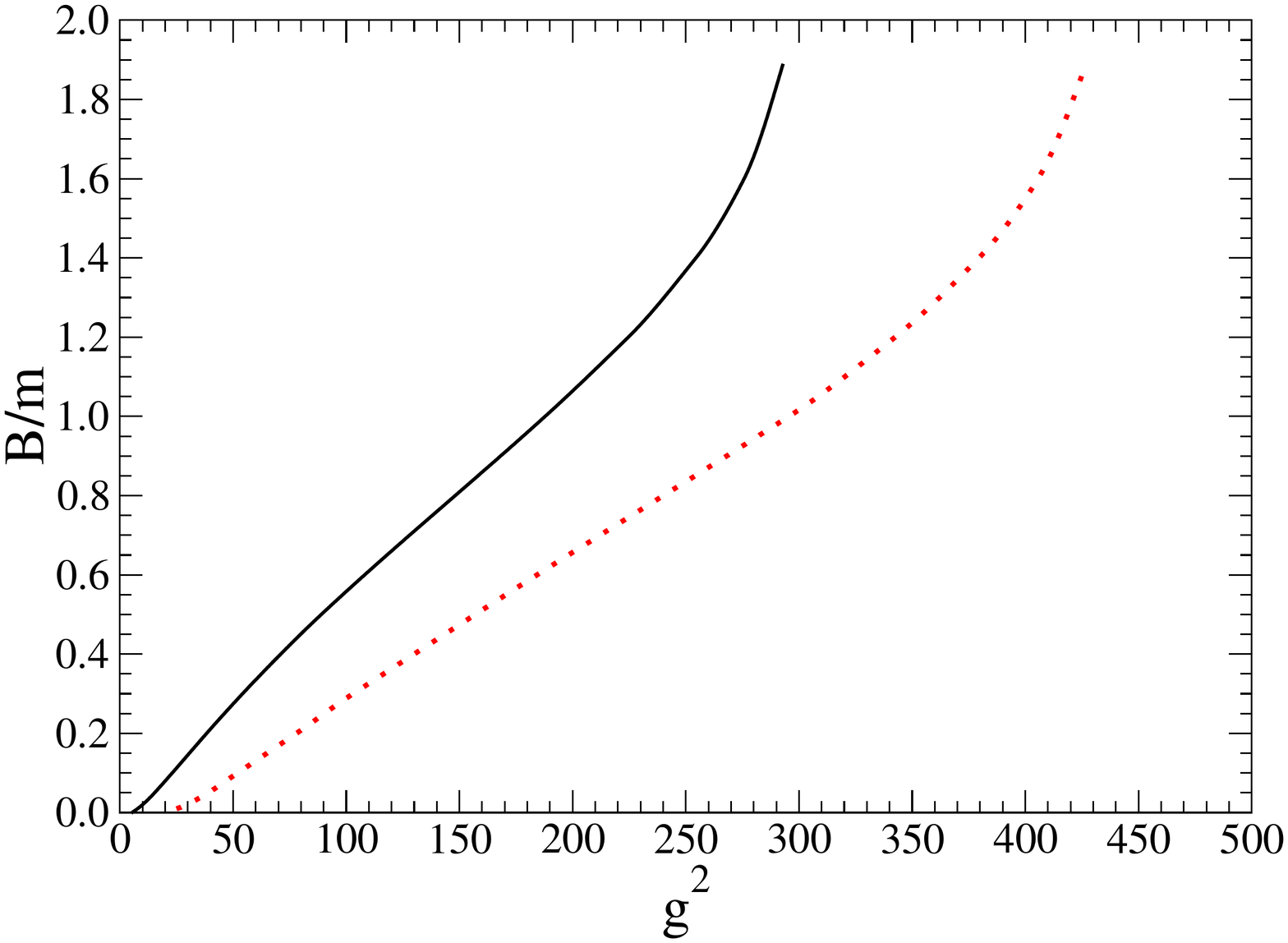}
   
   \includegraphics[width=9.0cm]{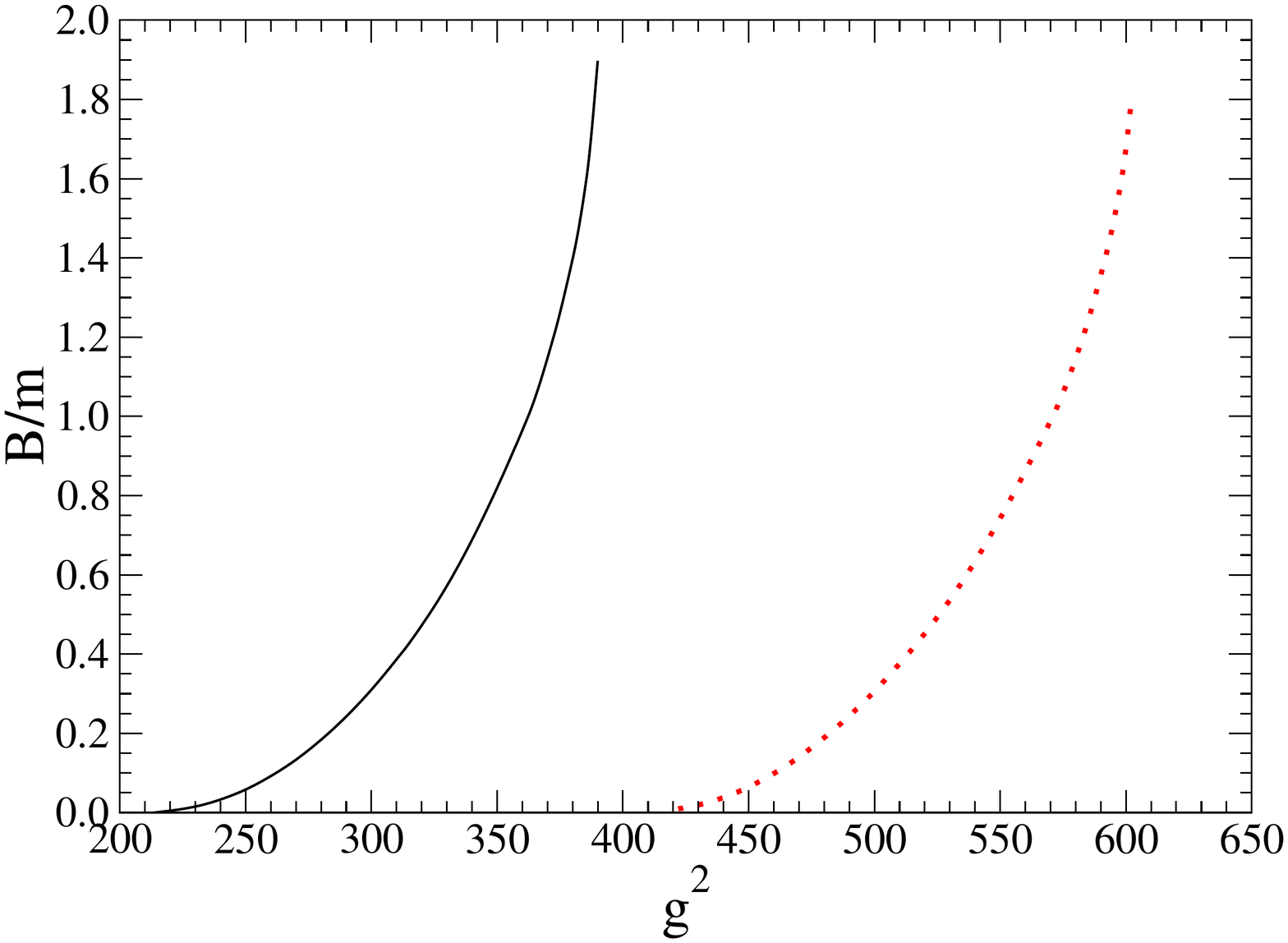}
   \caption{(color online) The binding energy $B/m$ vs $g^2$  for a
 scalar exchange (upper panel) and a pseudoscalar one (lower panel). {In both cases, the 
  vertex form-factor cutoff is $\Lambda/m=2$ (cf Eq. (\ref{vertexff})). 
  The masses of the 
 exchanged scalar/pseudoscalar boson are $\mu/m=0.15$ (solid line)  and $\mu/m=0.5$ (dotted line).}}
   \label{fig:1}
\end{figure} 

\begin{figure}[thb] 
\centering
   \includegraphics[width=9.0cm]{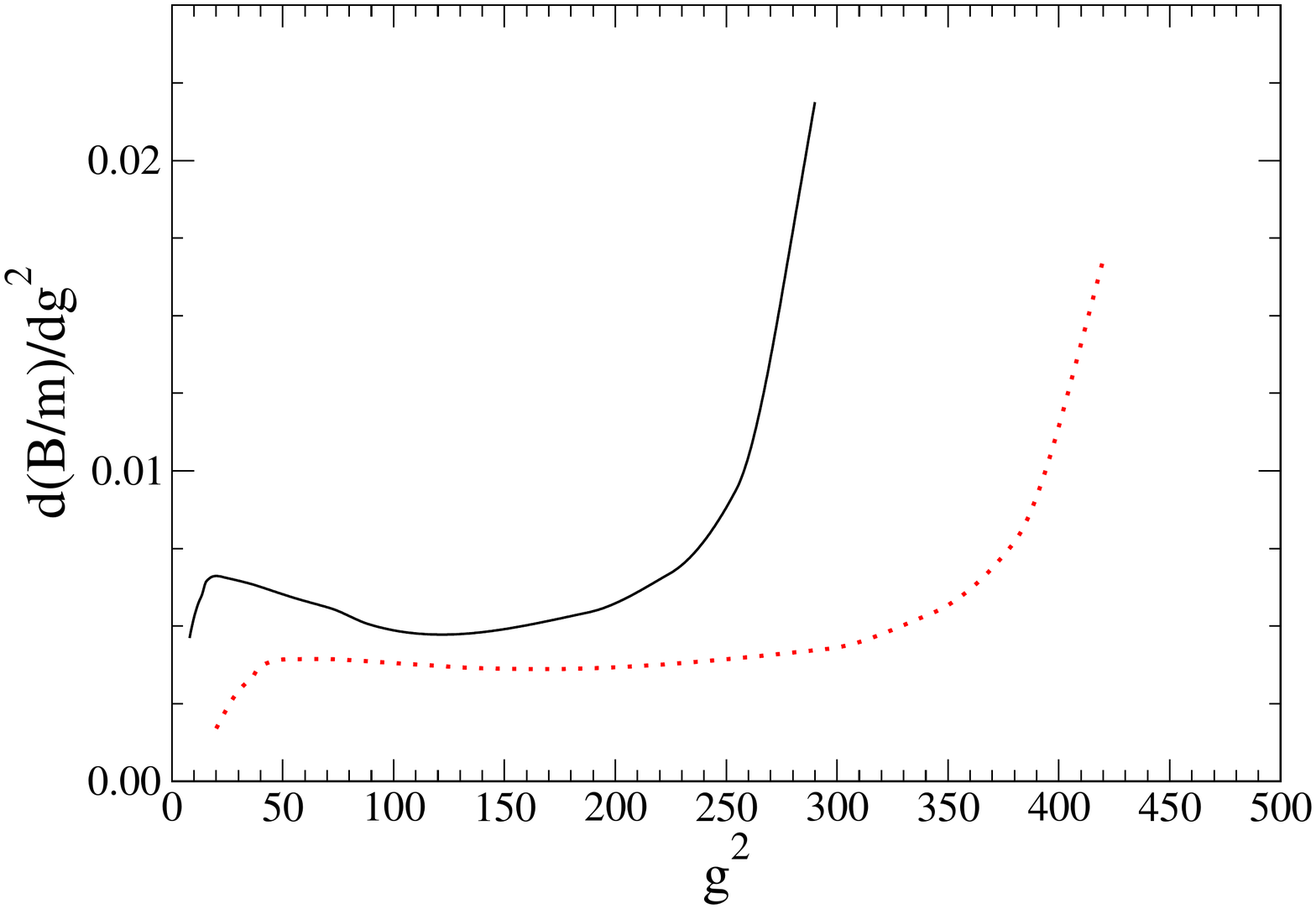}
   
   \includegraphics[width=9.0cm]{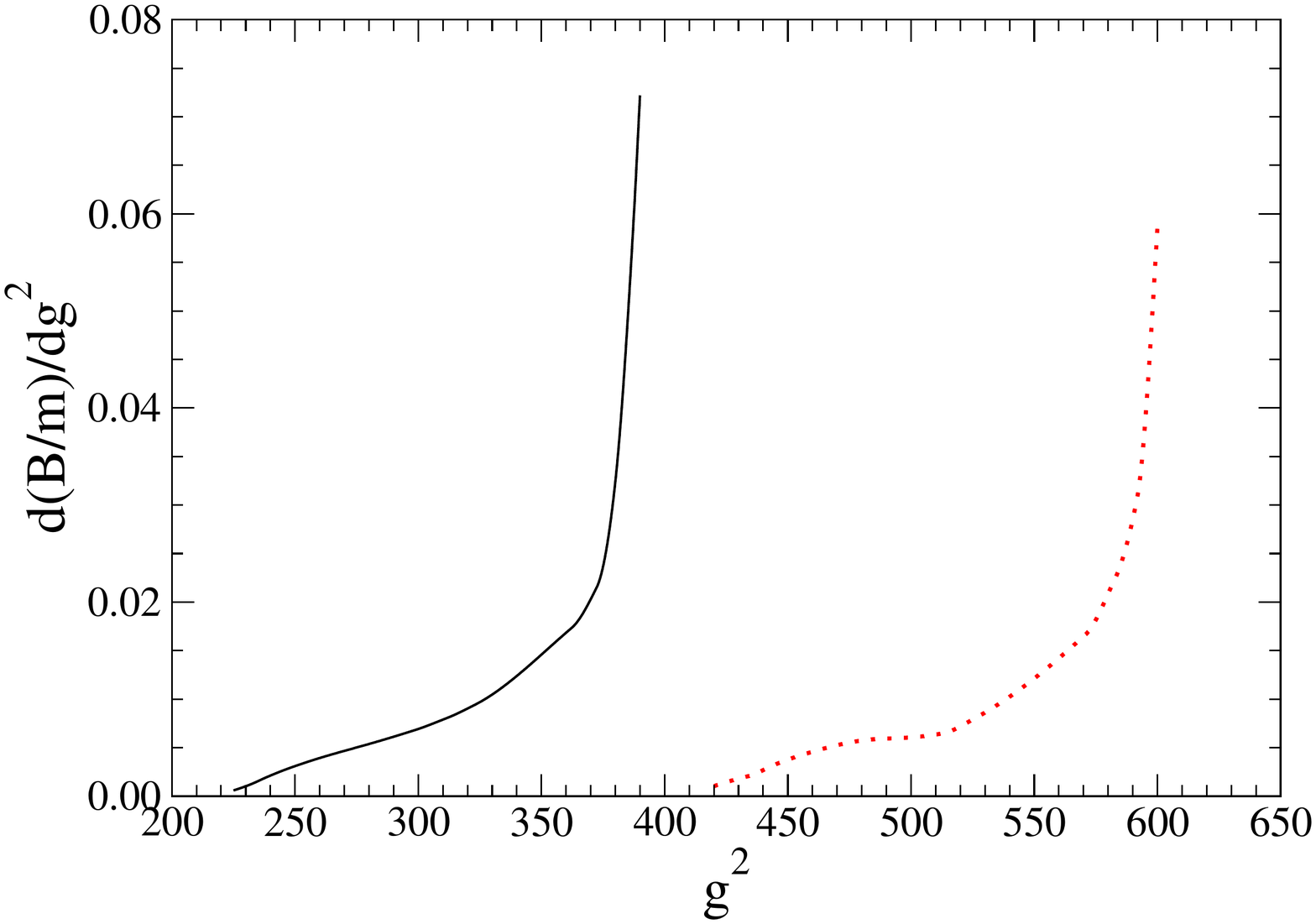}
   \caption{(color online) The same as in in Fig. \ref{fig:1}, but for the
   corresponding 
   derivatives $d(B/m)/dg^2$. }\label{fig:1b}
\end{figure} 

\subsection{Binding energy vs Coupling Constants}

We start by showing the binding energy $B/m$ as a function of  $g^2$  for 
both  scalar and pseudoscalar   {exchanges, fixing  the  cutoff in the 
vertex form factor,
 Eq. (\ref{vertexff}), to the value
 $\Lambda/m=2$. The results shown in Fig. \ref{fig:1}, partially 
  presented in Table I and II of Ref.
 \cite{dFSV1}, have been  obtained by  choosing the
  masses of the  scalar and 
 pseudoscalar exchanged bosons equal to  $\mu/m=0.15$ and $\mu/m=0.50$. In
 particular, the lines shown in   Fig. \ref{fig:1}  allow
 one to illustrate   interesting overall behaviors of
  the binding $B/m$, when the 
  boson mass $\mu$ changes. 
{ In general, for fixed values of $\Lambda/m$ and $B/m$, the coupling constant $g^2$ increases as the mass 
of the exchanged boson
increases. This feature is an expected one, if we take into account that 
 the interaction has a range  controlled by the
inverse of $\mu$. Hence, if the range  shrinks, then    an increasing 
of the interaction  strength is necessary for
 allocating a bound state with 
given $B/m$, as it is well-known also in the non relativistic case. 
In particular, for weakly bound systems, at fixed $B/m$,  the correlation
 between the coupling constant $g^2$ and $\mu/m$ is almost linear (cf 
  Tables I and II in Ref.
 \cite{dFSV1} for $B/m \le 0.05$).}
 
{  Moreover, the correlation between $g^2$ and $\mu$
is also driven by the nature of the exchanged boson, as shown by the 
quantitative differences illustrated in the upper
and the lower panels in Fig. \ref{fig:1}. As it is well-known in the  non
relativistic framework when the one-pion exchange is investigated, 
the pseudoscalar interaction 
is weak in a  $0^+$ state. In particular,  the tensor force obtained 
from the non relativistic reduction 
 does not contribute  in a $0^+$
 state, and the remaining interaction gives a repulsive contribution, making
 impossible to bind the system. Only when the relativistic effects become
 important, the binding can be established, but  for  very large $g^2$ . 
 In conclusion,   
 in the pseudoscalar case, one is 
  confronted with a much weaker attraction that requires, for a given binding
  energy, larger couplings 
  than  the scalar case needs.}

 When we approach   the collapse 
 of the bound state, i.e. when  a composite massless 
 particle is created since   $B/m \to 2$, the  shape of {the binding}
   becomes less sensitive to 
 the variation of {$g^2$ }  (a vertical  asymptote is approached).
  {By increasing the binding, the  size of the system becomes
    smaller and smaller and therefore }
  the range of the interaction,
 dictated by $\mu/m$, becomes less relevant {for the functional dependence of the
 binding upon $g^2$ close to its critical value for $B/m=2$,  namely  } the ultraviolet regime is now 
 starting to govern the dynamics inside
 the system. This holds for both exchanges.}
 
  Let us now focus on the  behavior of $d(B/m)/dg^2$, { shown in Fig.
 \ref{fig:1b}}. It is quite peculiar
  for the scalar case with
 respect to both pseudoscalar and vector cases 
 (for the figure illustrating $B/m$ for the massless 
 vector exchange see Ref. \cite{dFSV1}), since it shows the presence of
 a minimum {(less pronounced for $\mu/m=0.5$)} positioned almost in the middle of the range
 of $g^2$ relevant for a given $\mu$. {For the pseudoscalar case the minimum is close to the threshold,
  as a
 consequence of the repulsive contributions at low binding energies above mentioned. This observation suggests
 that also for the scalar case the appearance of a  minimum  be related to some
 repulsive contributions  that develop for $B/m\to 2$ (i.e. $M\to 0$). As a matter of fact, from a direct
 inspection of Table \ref{F0}, one can single out repulsive contributions that depends upon    $1/M$. }

\subsection{Light-front  amplitudes: Scalar case}

The LF amplitudes $\psi_i(\gamma,\,\xi;\kappa^2)$, Eq. (\ref{psii}), for the $0 ^+$ two-fermion system 
bound by a scalar exchange are presented in the Fig. \ref{fig:psiscgm0p15}.
 We { compare  two cases: weak binding
$B/m=0.1$ (upper panels) and strong binding $B/m=1.0$ (lower panels). The motivation of such a comparison 
is given by the attempt of  widening our 
intuition, based on non-relativistic physics, to the extreme binding, where the 
relativistic effects are expected to be relevant, like in the case of 
light pseudoscalar mesons. Indeed, in the present work 
an initial analysis of this 
 physical case will be 
proposed   by introducing  and
investigating a {\it mock pion}
  (see the next Sect. \ref{sect:mockpion}).
  
  \begin{figure}[thb] 
\centering
\includegraphics[width=8.5cm]{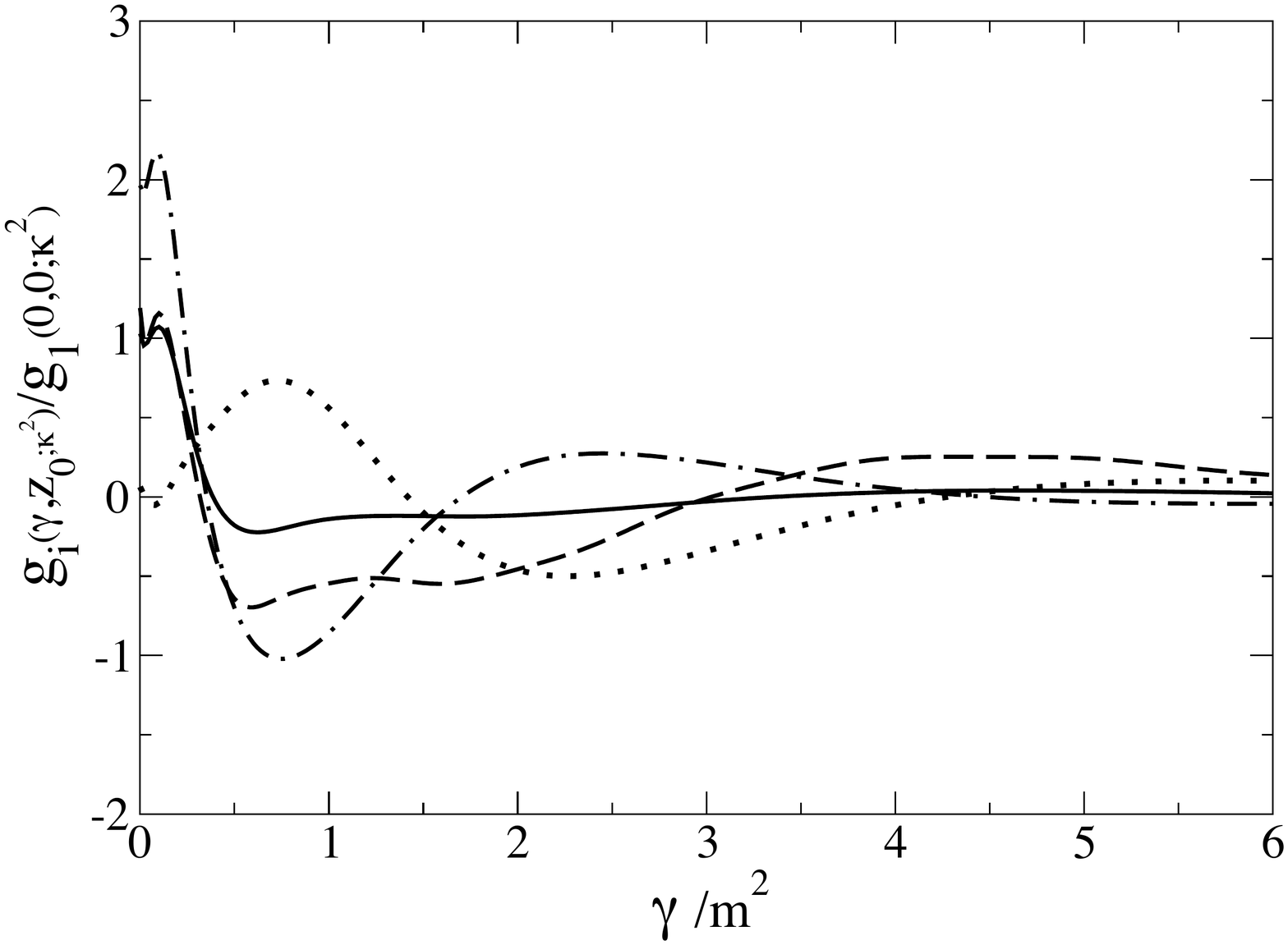}

\includegraphics[width=8.5cm]{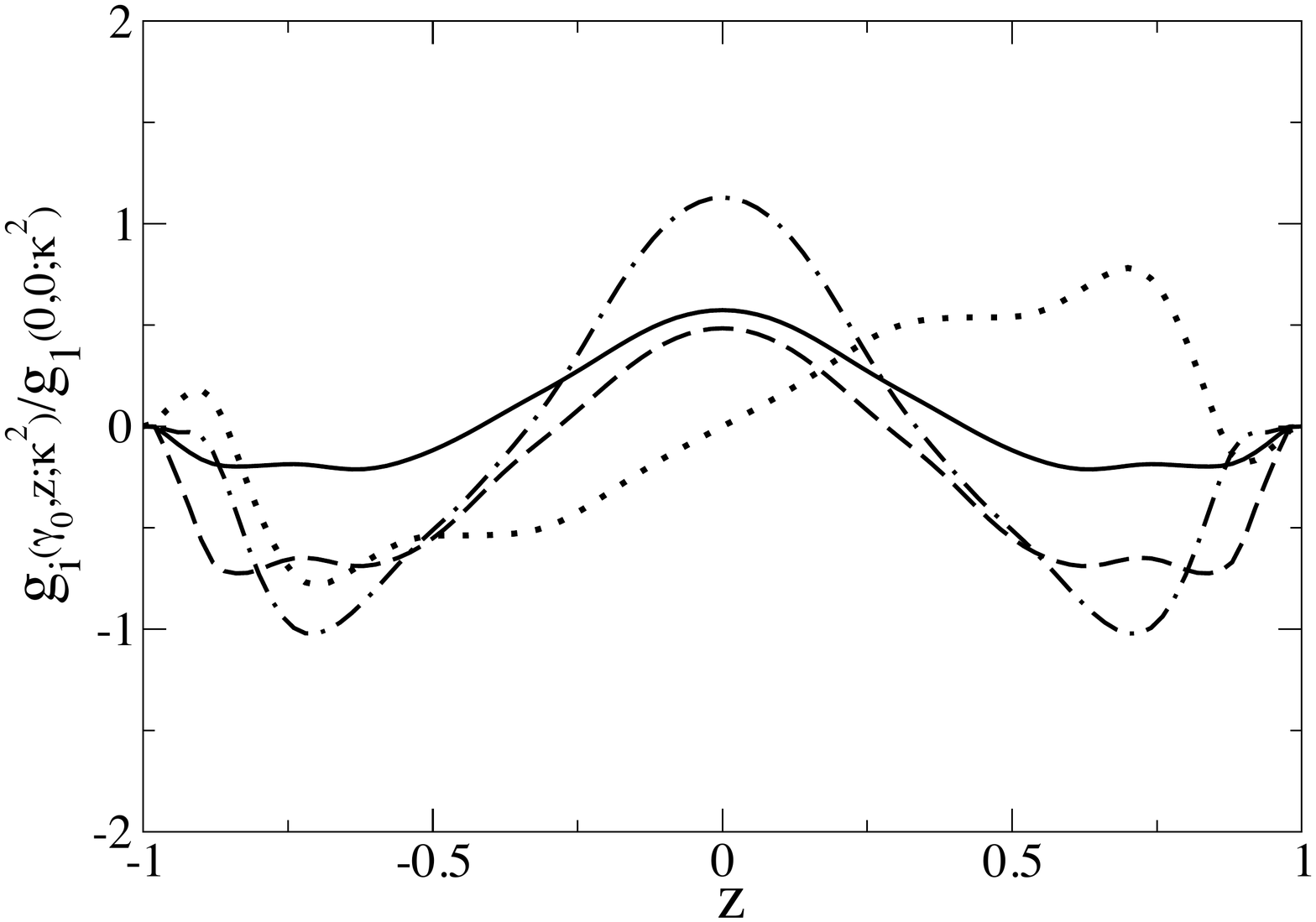}
\caption{ Nakanishi weight-functions  $g_i(\gamma,\,z;\kappa^2)$,  
   Eqs. (\ref{phinak}) and 
    (\ref{coupls1})
  evaluated for the $0^+$ two-fermion system 
   with a scalar boson exchange such that $\mu/m=0.5$ and $B/m=0.1$ (the corresponding coupling
   is $g^2=52.817$
   \cite{dFSV1}). The  vertex form-factor cutoff is 
   $\Lambda/m=2$. Upper panel:  $g_i(\gamma,\,z_0;\kappa^2)$ with $z_0=0.6$ 
   and running $\gamma/m^2$.
   Lower panel: $g_i(\gamma_0,\,z;\kappa^2)$ with $\gamma_0/m^2=0.54$ and
    running $z$, The 
   Nakanishi weight-functions  are normalized with respect to
   $g_1(0,0;\kappa^2)$.
    Solid line: $g_1$. Dashed line:  $g_2$. Dotted line: $g_3$. 
    Dot-dashed line: $g_4$.
    } 
   \label{fig:gnakscm0p5}
\end{figure}
 To begin, it is useful to compare our results for the Nakanishi weight-functions with the outcomes obtained in Ref. \cite{CK2010},
 within the Nakanishi framework,
but inserting a smoothing function.  
In Fig. \ref{fig:gnakscm0p5}, the amplitudes $g_i(\gamma,z;\kappa^2)$, 
corresponding to $\mu/m=0.5$ and $B/m=0.1$ (weak binding)
as in Ref. \cite{CK2010}, are presented. The  upper
panel shows $g_i(\gamma,z;\kappa^2)$ for a fixed values $z_0=0.6$ and running 
$\gamma$, while the
lower panel illustrates the same quantities, but for $\gamma_0/m^2=0.54$ and 
running $z$. Recall that
the results for
the eigenvalues obtained in Ref. \cite{CK2010}  coincide  with ours 
(cf \cite{dFSV1}) at the level of the
published digits, while 
 the eigenvectors, i.e. the Nakanishi weight-functions, are slightly 
 different. Indeed,
 it is extremely encouraging to observe that so much different   
methods for taking into account the singularities in the  BSE  are able 
to achieve an overall agreement,
(at least in weak-binding regime) in
describing $g_i(\gamma,z;\kappa^2)$,  that have   very rich   structures.

In Fig. \ref{fig:psiscgm0p15},  results obtained with $\Lambda/m=2$ and $\mu/m=0.15$ are presented.
 It should be pointed out 
  that the general behavior shown in  Fig. \ref{fig:psiscgm0p15} does not substantially change by 
 increasing the mass of the exchanged boson up to $\mu/m=0.5 $. Notice that the
 lhs of the figure contains $\psi_i(\gamma,\,\xi;\kappa^2)$ for a fixed value of $\xi$
  and running $\gamma$. The chosen value $\xi_0=0.2$ allows one to show in the
 same panel all the four $\psi_i$, since for $\xi=0.5$, where one should expect
 the maximal value (cf the rhs of the figure)  $\psi_3$ vanishes. All the LF
 amplitudes are normalized
 to $\psi_1(\gamma=0,\xi=0.5)$, to quickly appreciate the relative
 strengths (in a forthcoming paper \cite{dFSV2}, it will be adopted the 
  proper  normalization 
 through the 
 BS amplitude for  eventually obtaining the LF distributions).}

\begin{figure*}[thb] 
\centering
\includegraphics[width=7.0cm]{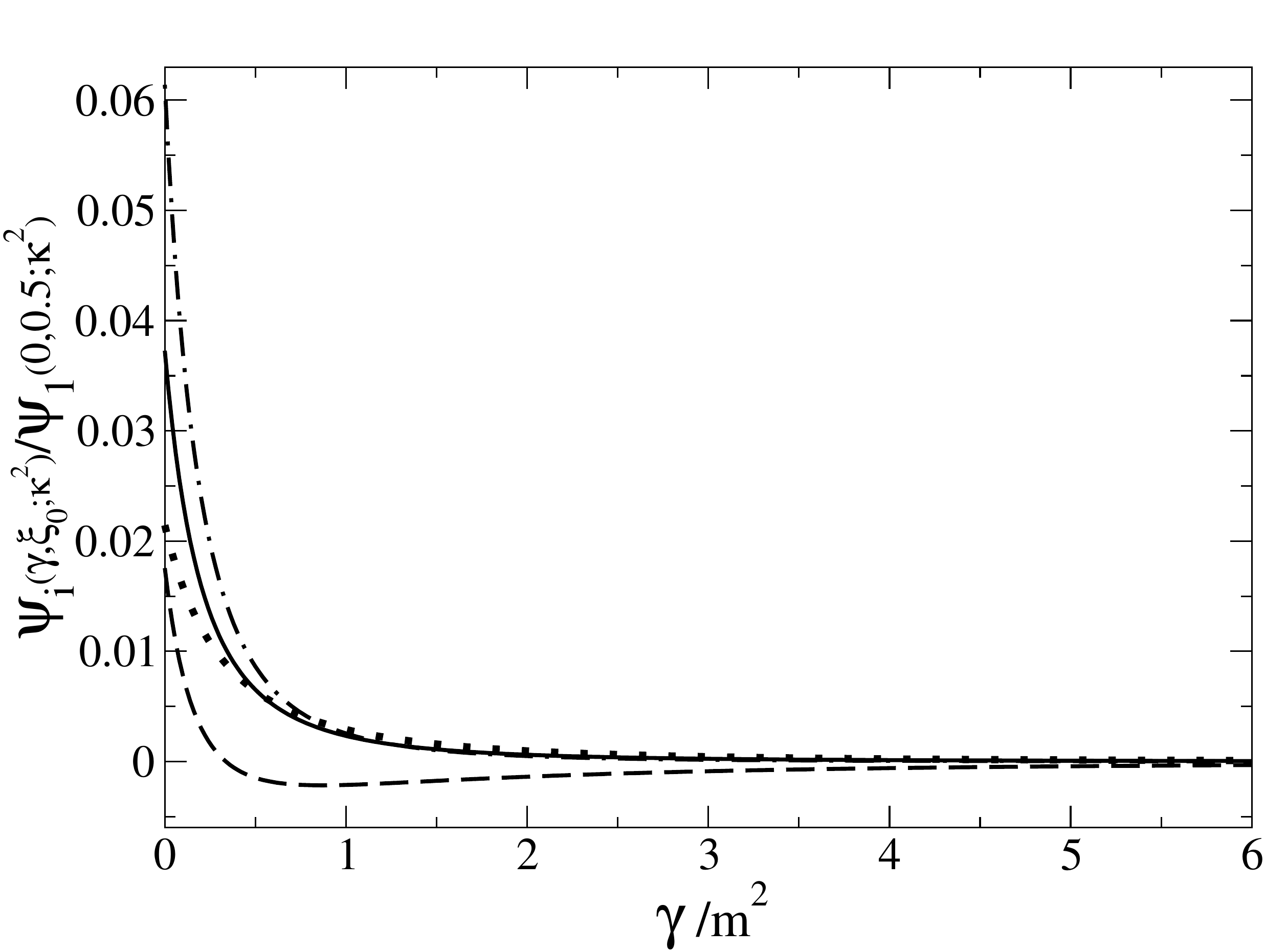}
\includegraphics[width=7.0cm]{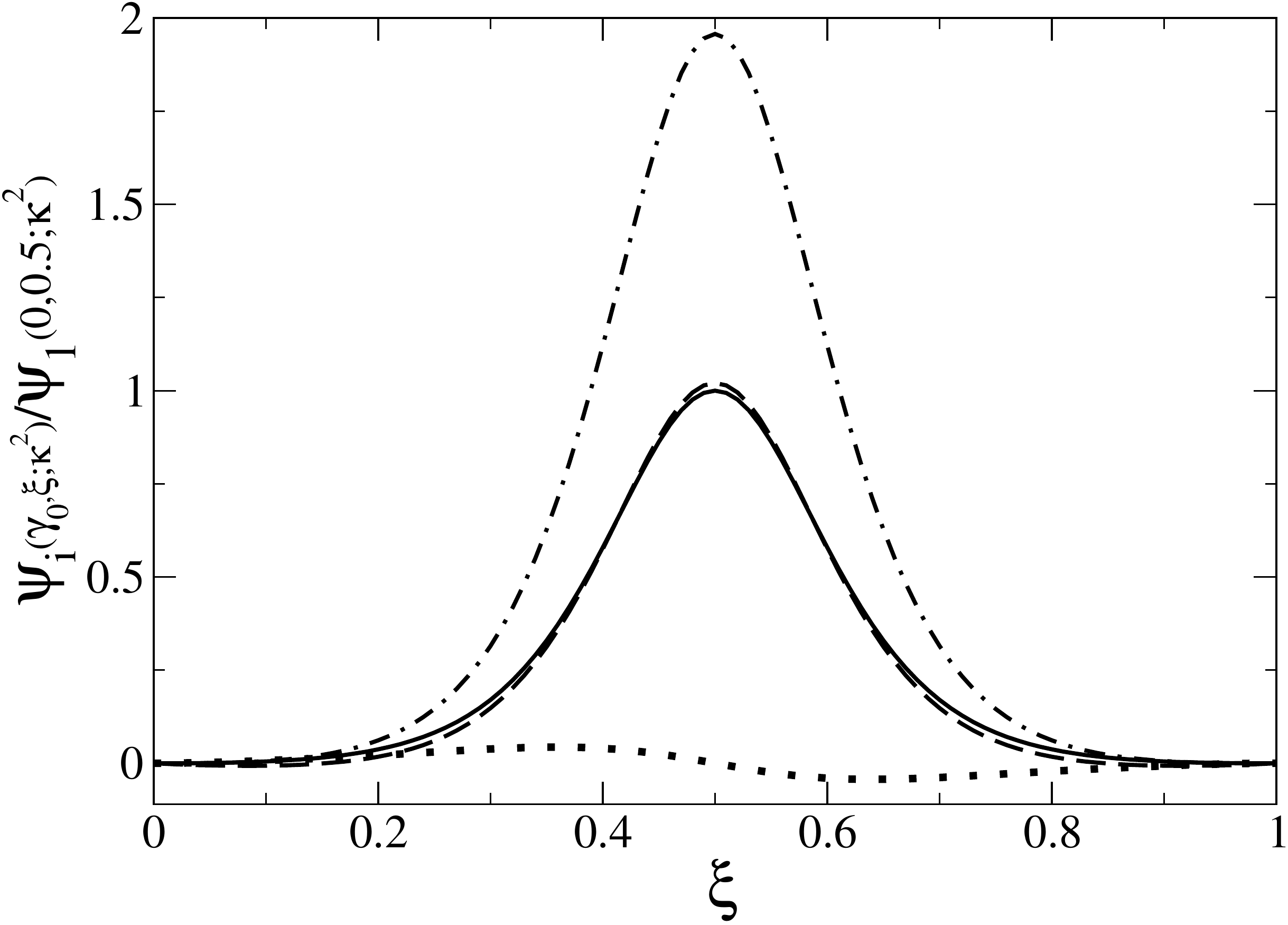}

\includegraphics[width=7.0cm]{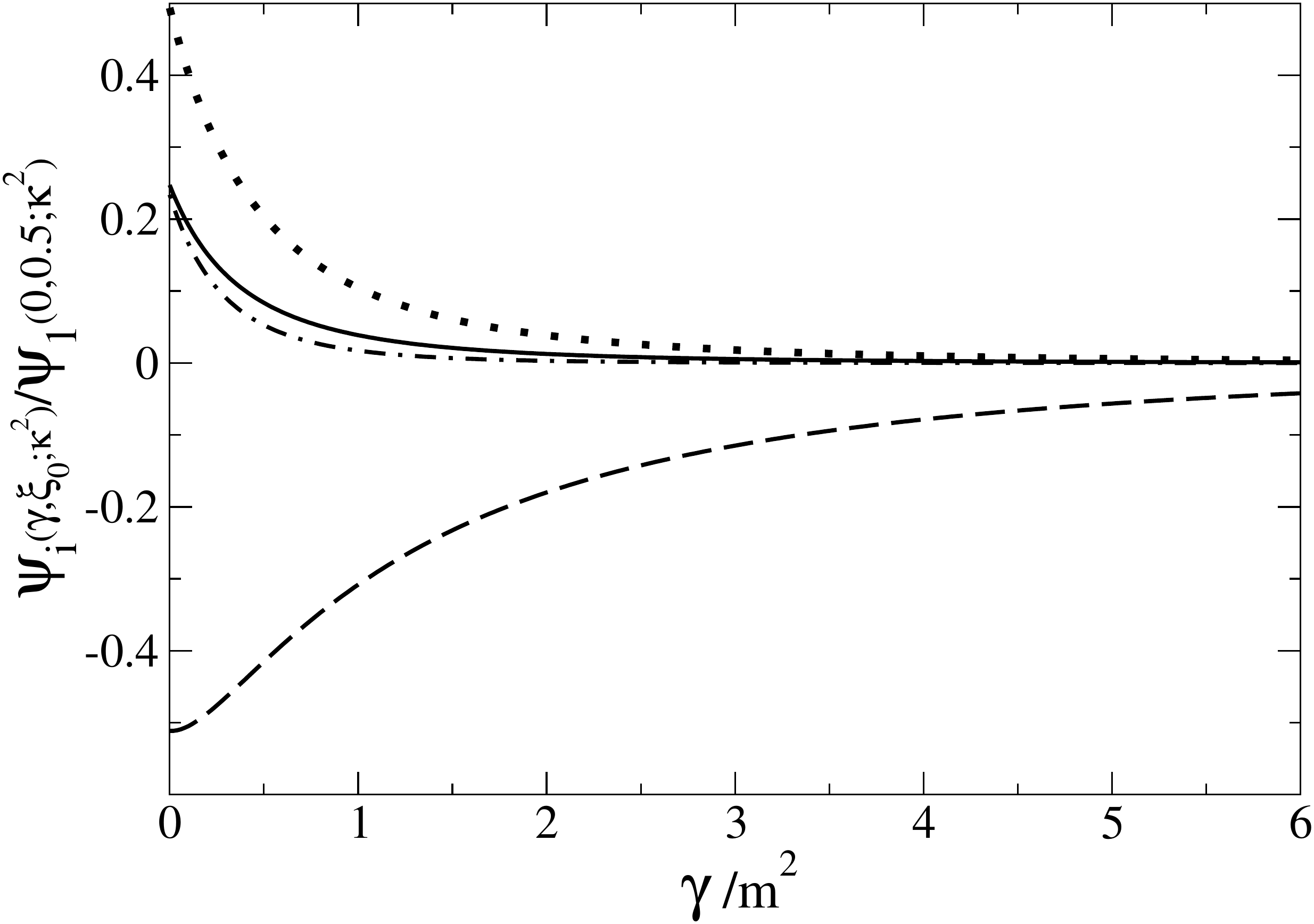}
\includegraphics[width=7.0cm]{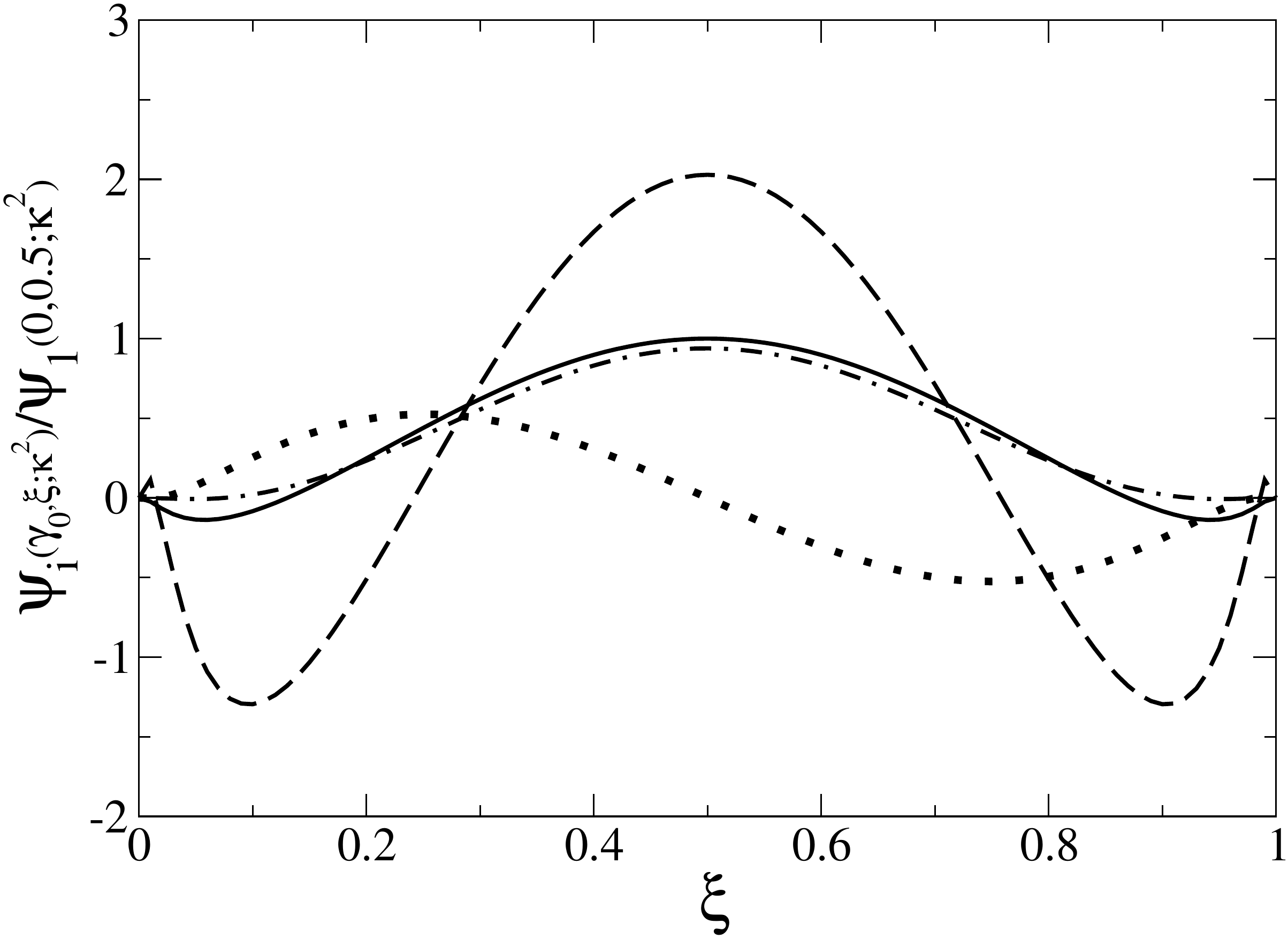}
   \caption{Light-front amplitudes  $\psi_i(\gamma,\,\xi;\kappa^2)$,  Eq. (\ref{psii}),
  evaluated for the $0^+$ two-fermion system 
   with a scalar boson exchange such that $\mu/m=0.15$. Upper  panel: weak binding  $B/m=0.1$ (the corresponding coupling
   is $g^2=23.12$
   \cite{dFSV1}). Lower  panel: strong binding
    $B/m=1.0$ (the corresponding coupling
   is $g^2=187.8)$
   \cite{dFSV1}). The  vertex form-factor cutoff is 
   $\Lambda/m=2$. N.B. $\gamma_0=0$  and $\xi_0=0.2$ (see text). The LF distributions are normalized respect to
   $\psi_1(0,\xi=0.50;\kappa^2)$.
    Solid line: $\psi_1$. Dashed line:  $\psi_2$. Dotted line: $\psi_3$. Dot-dashed line: $\psi_4$.} 
   \label{fig:psiscgm0p15}
\end{figure*}

 The behavior for running $\gamma$ shows a difference between the weak and strong
binding that can be directly ascribed to the size shrinking  of the
systems when the binding increases. Therefore the overall growing pattern of the 
amplitudes at higher $\gamma/m^2$ when $B/m$ increases (cf the
lhs of Fig. \ref{fig:psiscgm0p15} upper and lower panels)  is an expected one.
Remarkably, the characteristic momentum-width one can infer from the lines in Fig. \ref{fig:psiscgm0p15} 
 is of the order of { $\gamma/m^2\sim B/m$}. Another interesting feature is
  the growth of  
the amplitude 
$\psi_2$ when $B/m$ increases. This  reflects the importance of singular terms 
in the ultraviolet 
region, since in  this case one has the dominant role of the coefficient $c^S_{23}$ 
(see Eqs. (\ref{cij-1}) and (\ref{cij-2}) for counting the involved power of 
$k$).
Already in Ref. \cite{CK2010} it was noted the strong singular contribution 
of the kernel at the end points when 
$c_{23}$ is nonzero, which is enhanced in the relativistic case of 
strong binding.

The dependence on $\xi$ of the LF amplitudes $\psi_i$ 
presented  in rhs of Fig. \ref{fig:psiscgm0p15}
shows the expected peaks around 1/2, 
except for $\psi_3$, which is antisymmetric in $z$. Less trivial is the comparison
between the weak- and strong-binding regime, in particular for  $\psi_2$ and 
$\psi_4$. It can
be seen that for $B/m=1$, the LF amplitude $\psi_2$ accumulates 
toward the end points, a property which seems to be slightly present also
 in $\psi_1$. This feature can be traced to the same effect seen for running
 $\gamma$, namely the relevance of the coefficient $c^S_{23}$,  associated to
  the singular behavior at 
the end points. For the weak binding case, the effect is damped, 
as one could naively guess  from the observation that  
$ |\vec k|$ and $ |\vec k'|$ are of the order of $B\ll m$ (cf the upper left
panel) and  $c^S_{23}\sim B^3/M$, 
while for example $c^S_{21}=m\,M$.  Noteworthy $\psi_4$, 
 driving the spin-momentum correlation,  appears to be maximal for a weakly bound system
 for $\gamma=0$ in a state
$0^+$. The prominence of this component can be explained by considering the
first two coefficients, $c^S_{14}$ and $c^S_{24}$ 
(see Eqs. (\ref{cij-1}) and (\ref{cij-2})) that are proportional to 
$M^2$ and $M$
respectively. By comparing those coefficient with the corresponding $c^S_{11}$
and $c^S_{22}$ one can roughly understand the factor of two  between $\psi_4$
and $\psi_2$ at the peak $\xi=0.5$ for weak binding. The same relevance of the
$\psi_4$ component can be found for both the pseudoscalar and vector exchange (see
below). In particular, this last case leads us to focus the previous
discussion on the coefficients by excluding $c^S_{44}$, since $c^V_{44}=0$ but
the effect is present. Moreover, given the smallness of $\psi_3$ one can pay 
 attention  only  on $c^S_{14}$ and $c^S_{24}$.  

As a final remark, one should notice in the rhs that   the LF amplitudes
enlarge their-own range when $B/m$ grows, as expected from the size shrinking of
the system. 

\subsection{Light-front amplitudes: Pseudoscalar case}

 The general comments, with a particular emphasis on the relevance of the
coefficients producing LF singularities,  presented for the scalar case are 
suitable
 also for
the LF amplitudes  $\psi_i(\gamma,\,\xi;\kappa^2)$ obtained by using a pseudoscalar exchange with $\mu/m=0.15$.
But,  a crucial difference is generated by the peculiar dynamics entailed by 
 the spinor coupling characterizing the two cases, as already pointed out for 
 Fig. \ref{fig:1}. Such a difference plays a role in explaining the
 leading position of $\psi_4$, that is related to the
 spin-momentum correlations.

\begin{figure*}[htb] 
\centering
   \includegraphics[width=7.0cm]{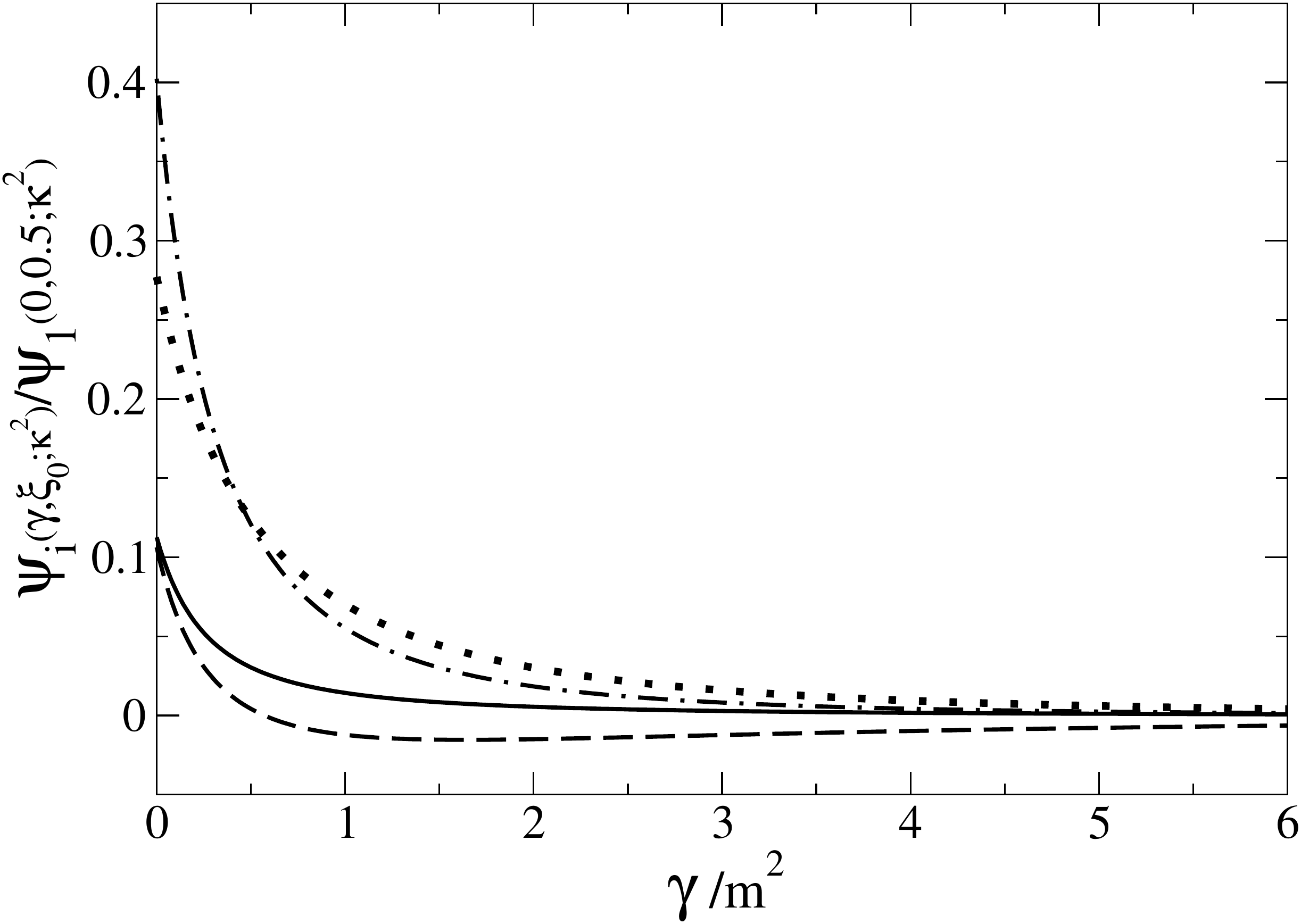}   
   \includegraphics[width=7.0cm]{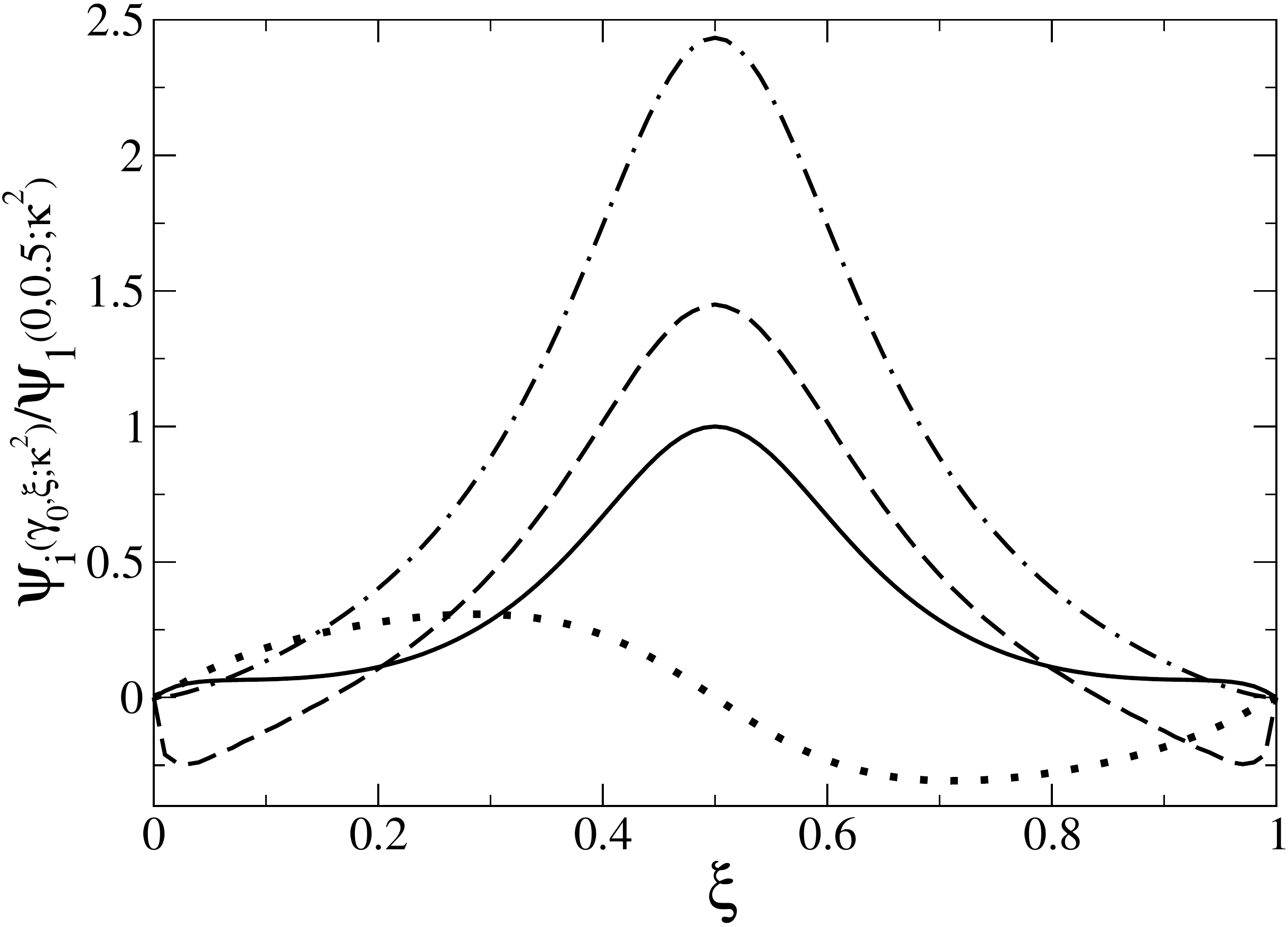}
   
   \includegraphics[width=7.0cm]{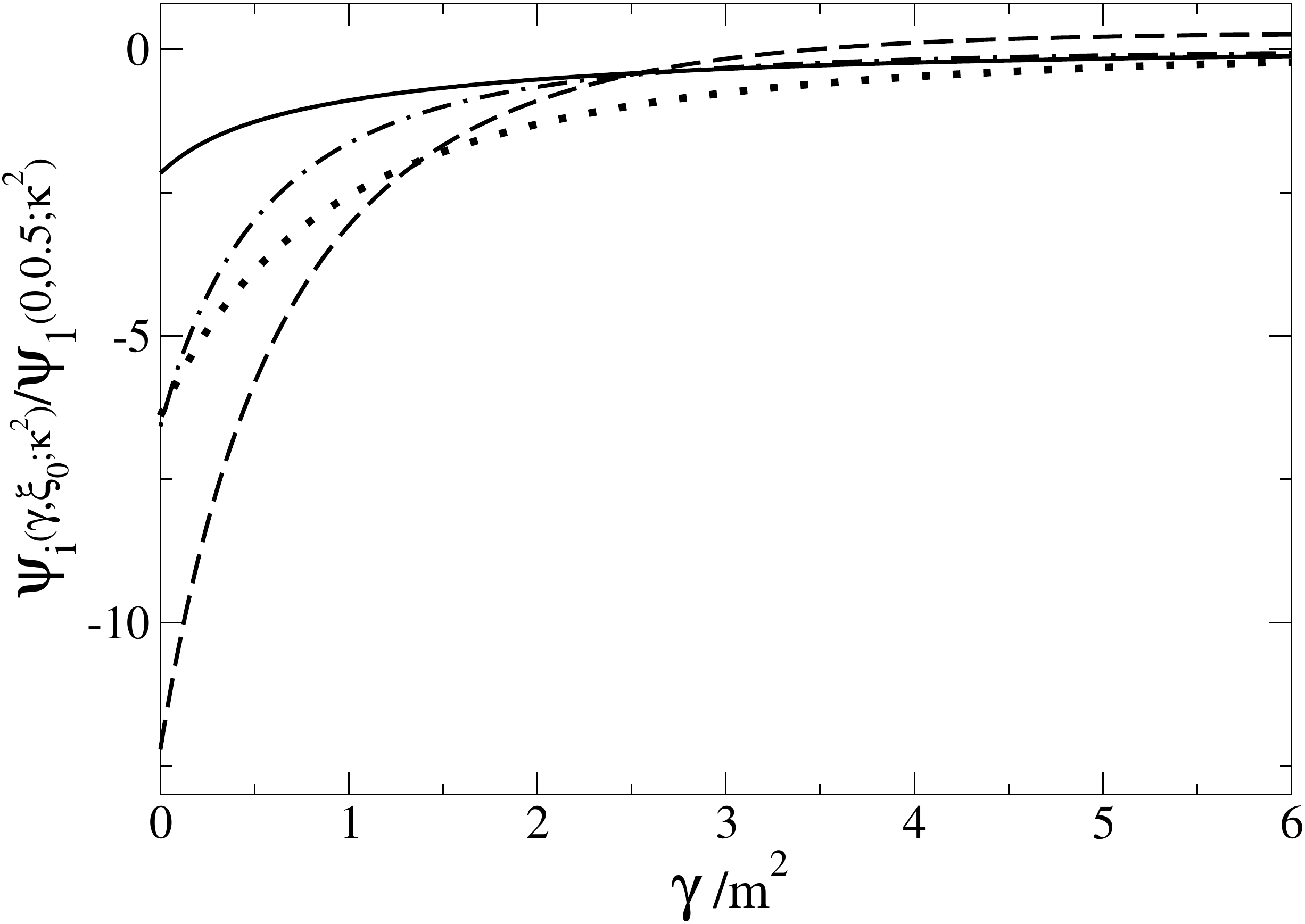}   
   \includegraphics[width=7.0cm]{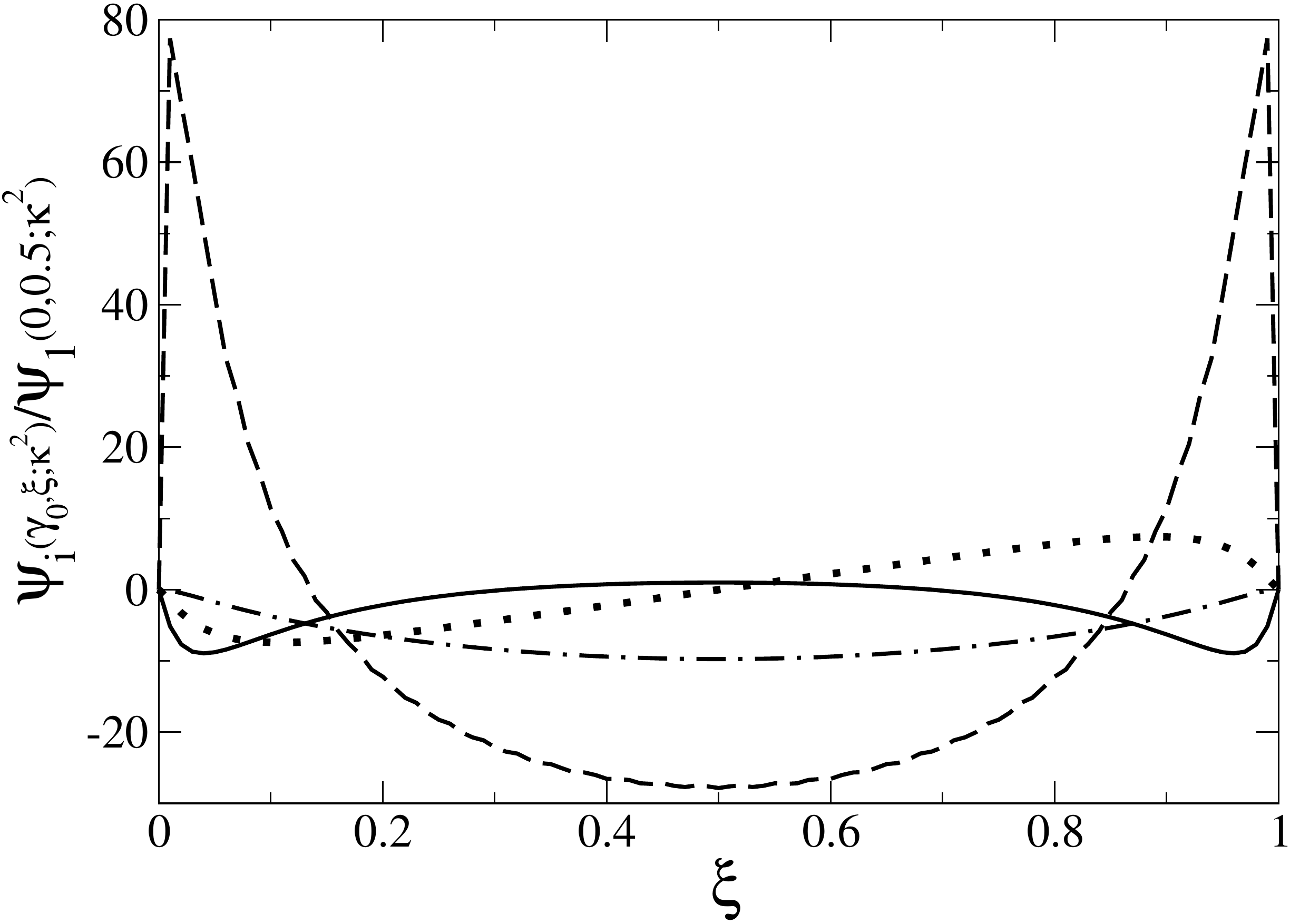}
\caption{Light-front amplitudes  $\psi_i(\gamma,\,\xi;\kappa^2)$,  Eq. (\ref{psii}),
  evaluated for the $0^+$ two-fermion system 
   with a pseudoscalar boson exchange such that $\mu/m=0.15$. Upper panels: weak binding  $B/m=0.1$ (the corresponding coupling
   is $g^2=262.1$
   \cite{dFSV1}). Lower panels: strong binding
    $B/m=1.0$ (the corresponding coupling
   is $g^2=362.3)$
   \cite{dFSV1}). The  vertex form-factor cutoff is
   $\Lambda/m=2$. N.B. $\gamma_0=0$  and $\xi_0=0.2$ (see text). 
   Solid line: $\psi_1$. Dashed line:  $\psi_2$. Dotted line: $\psi_3$. Dot-dashed line: $\psi_4$.} 
   \label{fig:psipscgm0p15}
\end{figure*}

The dependence of $\psi_i$ on $\xi$   is presented in the right panels of Fig. \ref{fig:psipscgm0p15}. 
For the weakly bound case, one notices that $\psi_1$ and $\psi_2$ are 
somewhat different while in the scalar case
they almost coincide. We trace back the reason for that, by 
 looking at the spinor structure associated to each 
$\psi_i$. The Dirac operator for $\psi_1$ is $\gamma_5$ and for $\psi_2$ is 
$\psla{p}\gamma_5/M$. For both cases
in the weak binding limit, namely in the non-relativistic regime,  the vector 
charge and scalar 
charge densities of the state are expected to be close, and in the scalar coupling case the 
corresponding operators commutes with the scalar coupling Dirac operator. 
For the pseudoscalar case, 
the $\gamma_5$ coupling has opposite commutation properties with the Dirac 
structure associated to $\psi_1$ and
$\psi_2$, and this feature  can be identified as the source for the observed 
difference in the upper-left panel of Fig. \ref{fig:psipscgm0p15}.

In conclusion, the performed analysis of scalar and pseudoscalar exchanges
yields a description coherent with the intuition stemming from the general structure of both
the BS amplitude and the kernel, and increases the confidence in the novel
approach we are pursuing for solving BSE in { 
Minkowski space}. Therefore, to
address the
vector exchange with such a reliable tool becomes a very stimulating issue, 
given its possible application to hadron physics.

\subsection{Light-front  amplitudes:  Vector case}

Starting  from the solution of the  four-dimensional BSE,
our goal is to construct a phenomenological tool for  analyzing
the structure  of a simple model of the pion, namely  a quark-antiquark pair 
living in {Minkowski} space and bound through a massive vector  exchange. 

\begin{table}
 \caption{The  
 vector coupling vs the binding energy for a massless vector exchange, i.e. $\mu/m=0$. 
 First column: binding energy.
 Second column: coupling constant $g^2$,  obtained
 by taking analytically into account the fermionic singularities, (see text).
 Third column:  results obtained in  Ref. \cite{CK2010} where the 
 singularities are
 treated  numerically, by using a smoothing function. 
 The vertex form factor cutoff is $\Lambda/m=2$.} \label{tabvec}
 \begin{center}
 \begin{tabular}{|c|c|c|}
 \hline
 B/m&$g^2_{dFSV}$(full)  &~~~$g^2_{CK}$~~~\\
 \hline
 0.01& 3.273   &3.265\\
 0.02& 4.913 & 4.910\\
 0.03& 6.261 & 6.263\\
 0.04& 7.454 & 7.457\\
 0.05& 8.548 & 8.548\\
 0.10& 13.15 & 13.15\\
 0.20& 20.43 &20.43\\
 0.30& 26.51 &26.50\\
 0.40& 31.86 &31.84\\
 0.50& 36.66 &36.62\\
 1.00& 54.62 & - \\
 1.20& 59.51 & -\\
 1.40& 63.23 & - \\
 1.60& 65.86 & -\\
 1.80& 67.43 & -\\
 \hline
 \end{tabular}
 \end{center}
 \end{table}

Before considering a quark-antiquark system, we continue  our analysis of the 
changes in the structure of the $0^+$ state with 
different couplings and binding energies in order to fully appreciate the 
subtleties generated  by the
dynamics inside the bound state. It is worth bearing in mind how accurate are
our calculations by  presenting a quantitative comparison with analogous
calculations shown in Ref. \cite{CK2010}. The massless exchange
between  two fermions in the $0^+$ state was considered  { (i.e. Eq. (\ref{kernv}) with $\mu^2=0$)}. As shown in Table
\ref{tabvec}, the agreement is excellent, and moreover the possibility to have a
formal treatment of the LF singularities allows us to extend the range of the
calculations. The behavior of  $B/m$ as a function of $g^2$ (see the
corresponding figure in \cite{dFSV1})
has the same overall structure found for  the pseudoscalar case, 
with a minimum of the
derivative of $d(B/m)/dg^2$ close to the threshold.

Given that, before treating the {\it mock pion} problem, we present results 
for a massive vector with $\mu/m=0.15$, still using  
the form factor parameter equal to  $\Lambda/m=\, 2$. As it is easily seen, in
Fig. \ref{fig:psivecgm0p15}  general patterns similar to the ones observed for
the  pseudoscalar
case can be recognized, when one moves from  weak- to strong-binding regime.
\begin{figure*}[htb] 
\centering
   \includegraphics[width=7.0cm]{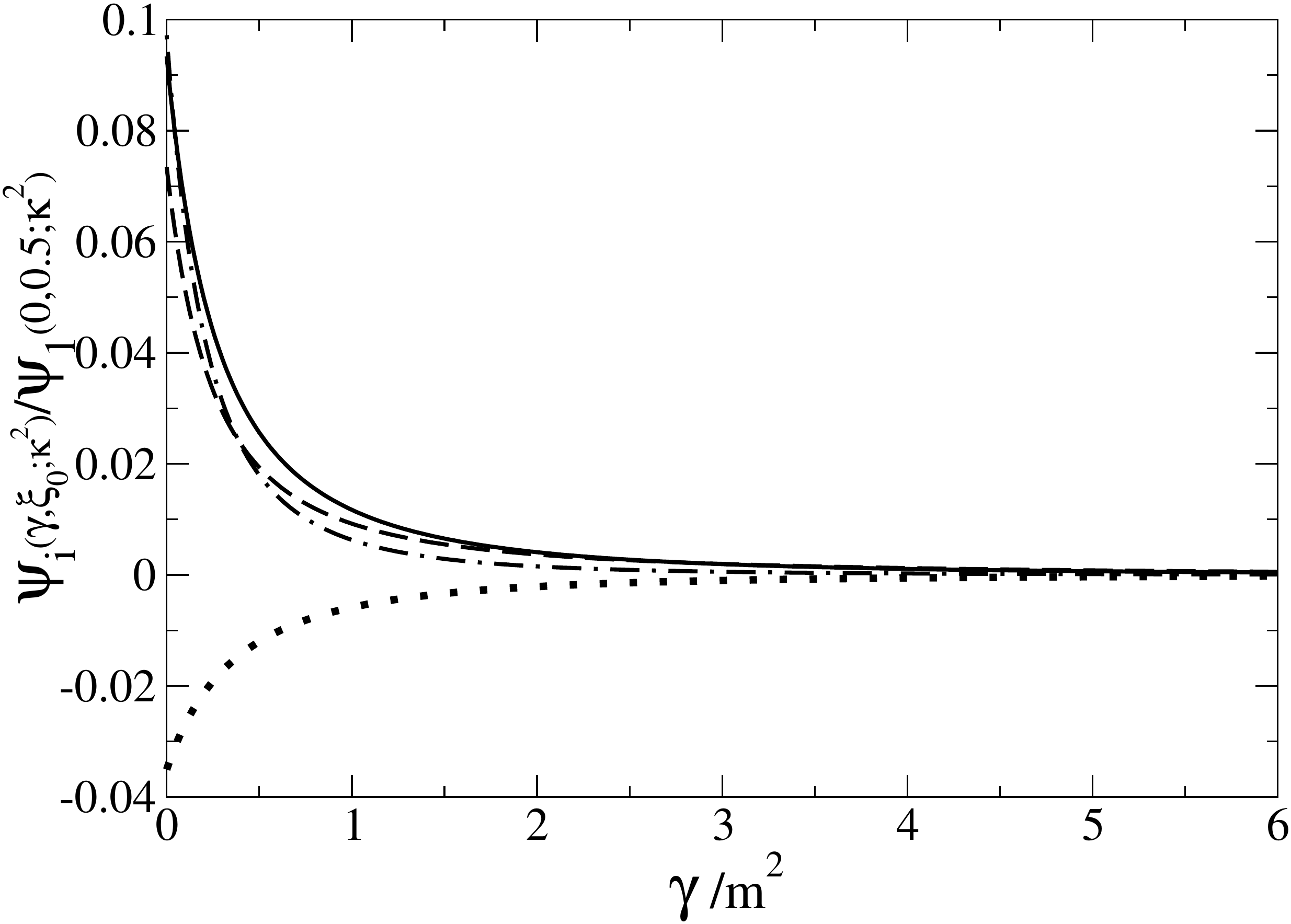}   
   \includegraphics[width=7.0cm]{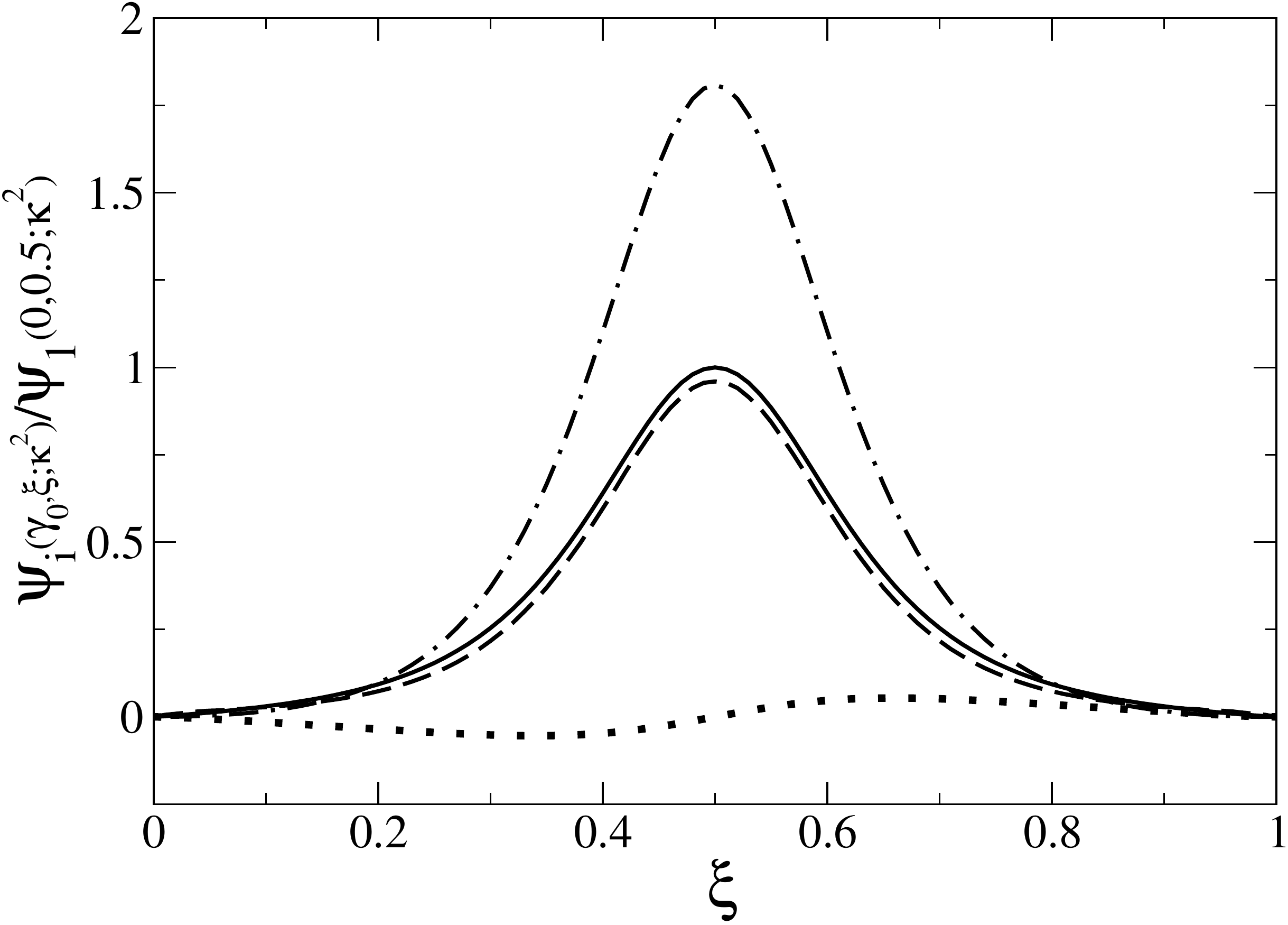}
     
   \includegraphics[width=7.0cm]{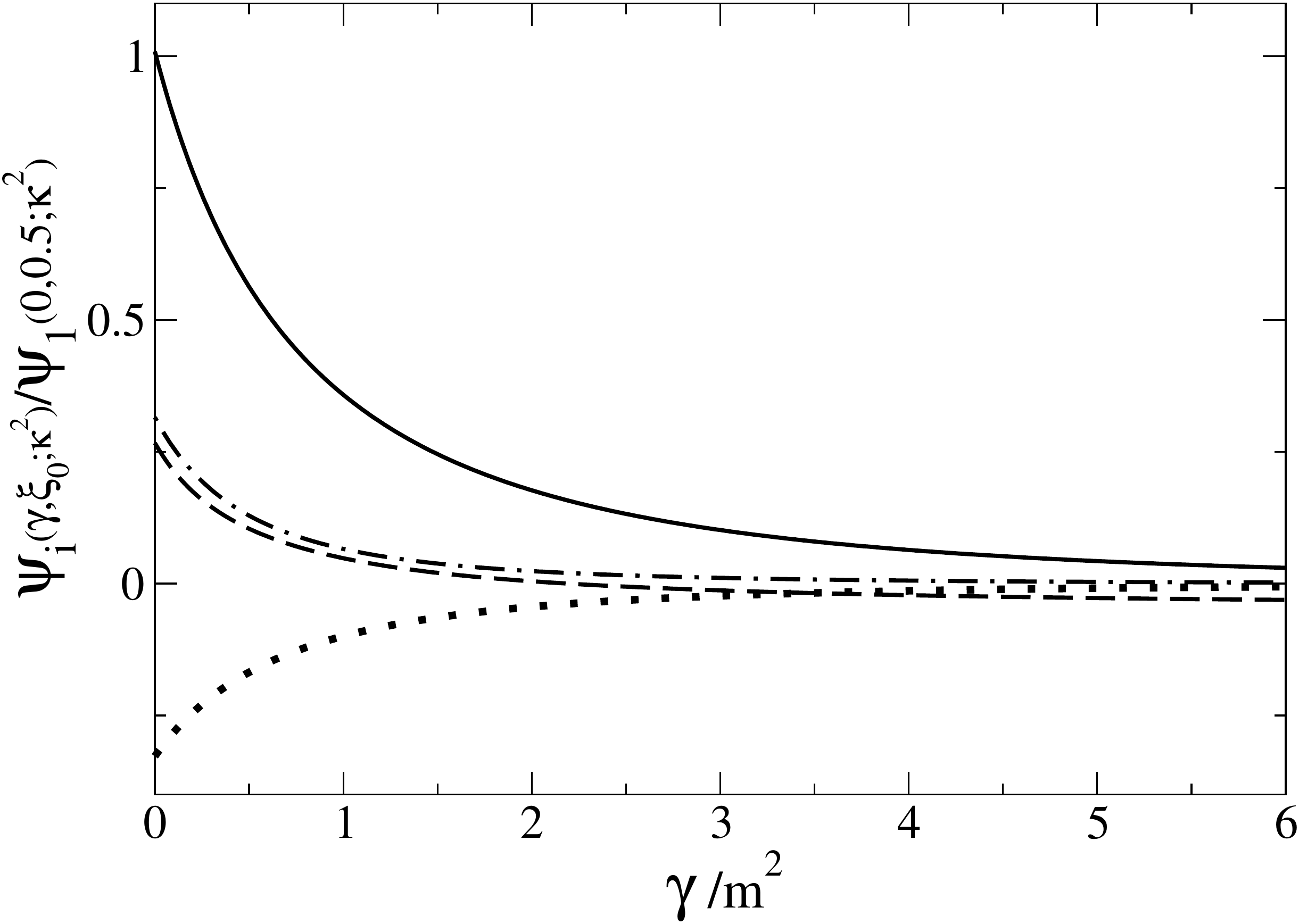}   
    \includegraphics[width=7.0cm]{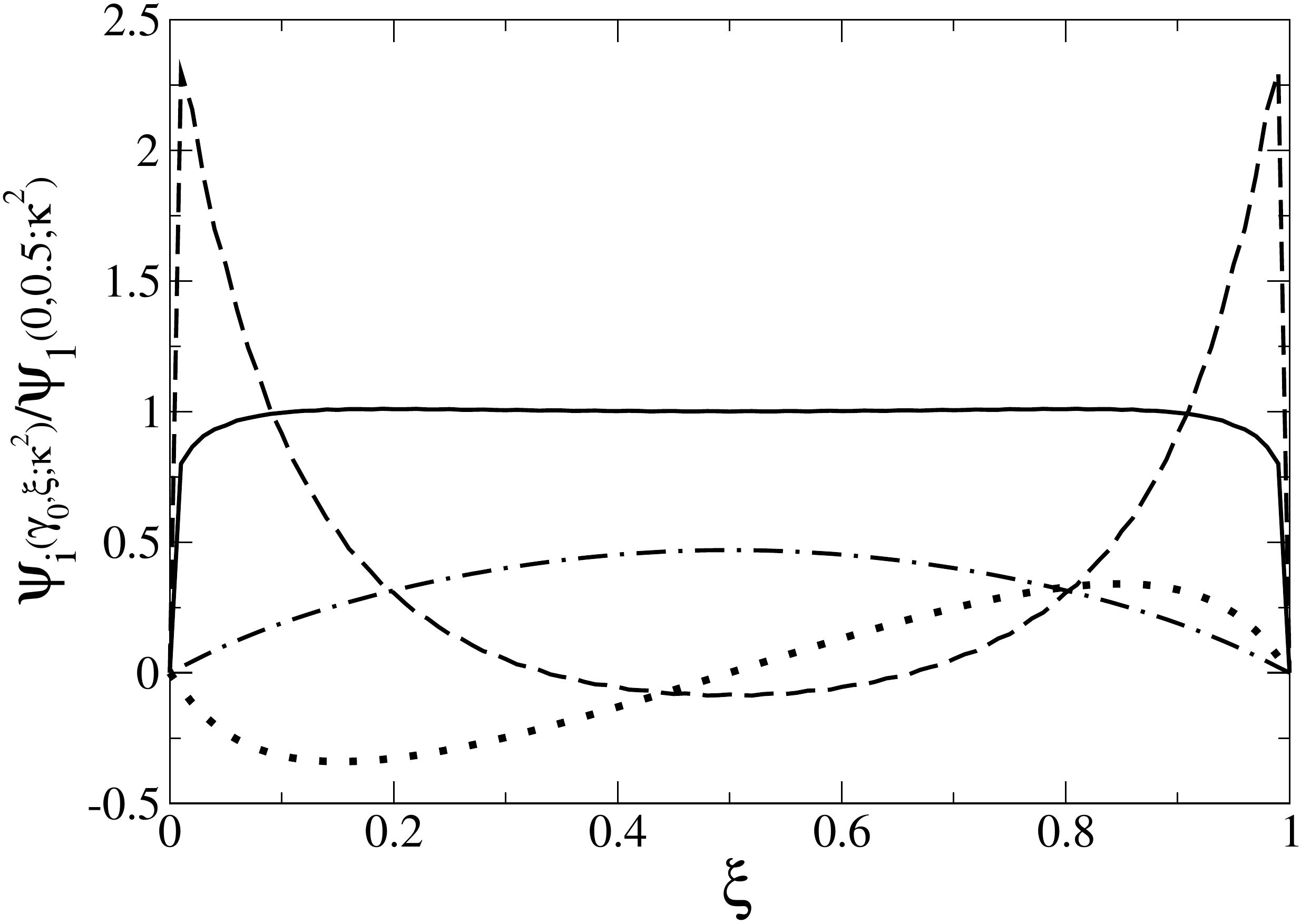}
   \caption{Light-front amplitudes  $\psi_i(\gamma,\,\xi;\kappa^2)$,  Eq. (\ref{psii}),
  evaluated for the $0^+$ two-fermion system 
   with a massive vector exchange such that $\mu/m=0.15$. Upper  panel: weak binding  $B/m=0.1$ (the corresponding coupling
   is $g^2=16.65$). Lower   panel: strong binding
    $B/m=1.0$ (the corresponding coupling
   is $g^2=59.23)$). The  vertex form-factor cutoff is
   $\Lambda/m=2$. N.B. $\gamma_0=0$   and $\xi_0=0.2$ (see text). 
   Solid line: $\psi_1$. Dashed line:  $\psi_2$. Dotted line: $\psi_3$. Dot-dashed line: $\psi_4$.} 
   \label{fig:psivecgm0p15}
\end{figure*} 
 The dependence of the LF amplitudes $\psi_1$ and $\psi_2$ 
on $\xi$,  in the right panels of the 
figures,  suffers a  dramatic change  from weak to strong binding, while 
 the amplitudes 
$\psi_3$ and $\psi_4$ just get wider when the binding increases  (apart the
change of sign of $\psi_3$). As in the scalar case, the 
amplitudes  $\psi_1$ and $\psi_2$ almost coincide for running $\xi$ and  $\gamma=0$,
 in the weak binding case.  This feature  
can  be understood by considering the Dirac structure
 associated with $\psi_1$ and $\psi_2$ in the expansion of the BS amplitude, 
 { i.e. $\gamma_5$ for the first
 amplitude and
 $\gamma_0\gamma_5$ for the second one, in the CM frame. In the non relativistic limit, the vector
 interaction largely reduces to its Coulomb component,  whose Dirac structure, $\gamma_0$, has 
 equal
 commutation properties with the ones associated to $\psi_1$ and $\psi_2$. The same happens when the
 scalar exchange is considered, since one has simply the identity matrix at the interaction vertex.
 Moreover,
  in the weak binding regime, the scalar and charge densities of the fermionic constituents (acting at the
  interaction vertices) tend to be the same, while for growing $B/m$ this is not the case.
  Finally, notice that even the coupling 
constants $g^2$ are similar for scalar and vector exchange, in the weak binding regime.}

Moving  from  weak to  strong binding the amplitudes $\psi_1$, $\psi_3$ and 
$\psi_4$ become wider as a function of $\xi$. A new feature arises 
in $\psi_1$, it becomes quite flat and sharply decreasing at the end points, 
while $\psi_2$ has the characteristic peaks discussed before. 
The flattening of $\psi_1$ appears to be a signature of the vector coupling, 
since it remains present  when the vector mass increases, as  illustrated in the
next Section.  Indeed, all the patterns shown in Fig. \ref{fig:psivecgm0p15} 
do not  qualitatively change 
 when the mass of the exchanged boson grows.

\section{The mock pion}
\label{sect:mockpion}
 {In this Section, we present a first  investigation of
 a simplified model for the pion, that, at some extent, can be  considered  
  Lattice-QCD 
 inspired. 
 In particular, we have tuned  our
  parameters, like the masses of  quark 
 and  
 exchanged vector boson,   according 
 to the outcomes of  Lattice-QCD calculations.

 As  above illustrated, within  the formal elaboration one adopts 
for getting solutions of BSE, it is natural to switch to a BS amplitude
pertaining to a fermion\hyp{}antifermion system. If one  multiplies the original fermion-fermion BS amplitude by the
charge-conjugation operator, then   one  gets 
 a compact expression of the BSE with fermionic degrees of freedom, eventually using the standard
multiplication  rule for matrices. Summarizing, the scalar functions (Eq. (\ref{bsa})) that enter
both fermion-fermion ($0^+$) and fermion\hyp{}antifermion ($0^-$) BS amplitudes
 are the same, but the Dirac
structures are different,  as it must be. The ladder  kernel we 
exploit is the one corresponding to a massive vector
exchange, in Feynman gauge (cf Eq. (\ref{kernv})), 
with the same  vertex form factor  in Eq. (\ref{vertexff}).

\begin{table}[t]
 \caption{Coupling constants for our Lattice-QCD inspired  {\em mock pion},
 obtained  with $B/m=1.44$ and
 $\mu/m=2.0$. 
   Two values for vertex form factor cutoff  $\Lambda/m$, have been chosen.} 
  \label{tab2}
 \begin{center}
 \begin{tabular}{|c|c|c|}
 \hline
 $\Lambda/m$ &$g^2$ &$\alpha_s$ Eq. (\ref{alphas})\\
 \hline
 3& 435.0   &10.68\\
 8& 52.0 & 3.71\\
 \hline
 \end{tabular}
 \end{center}
 \end{table}

 \begin{figure*}[htb] 
\centering
    \includegraphics[width=8.0cm]{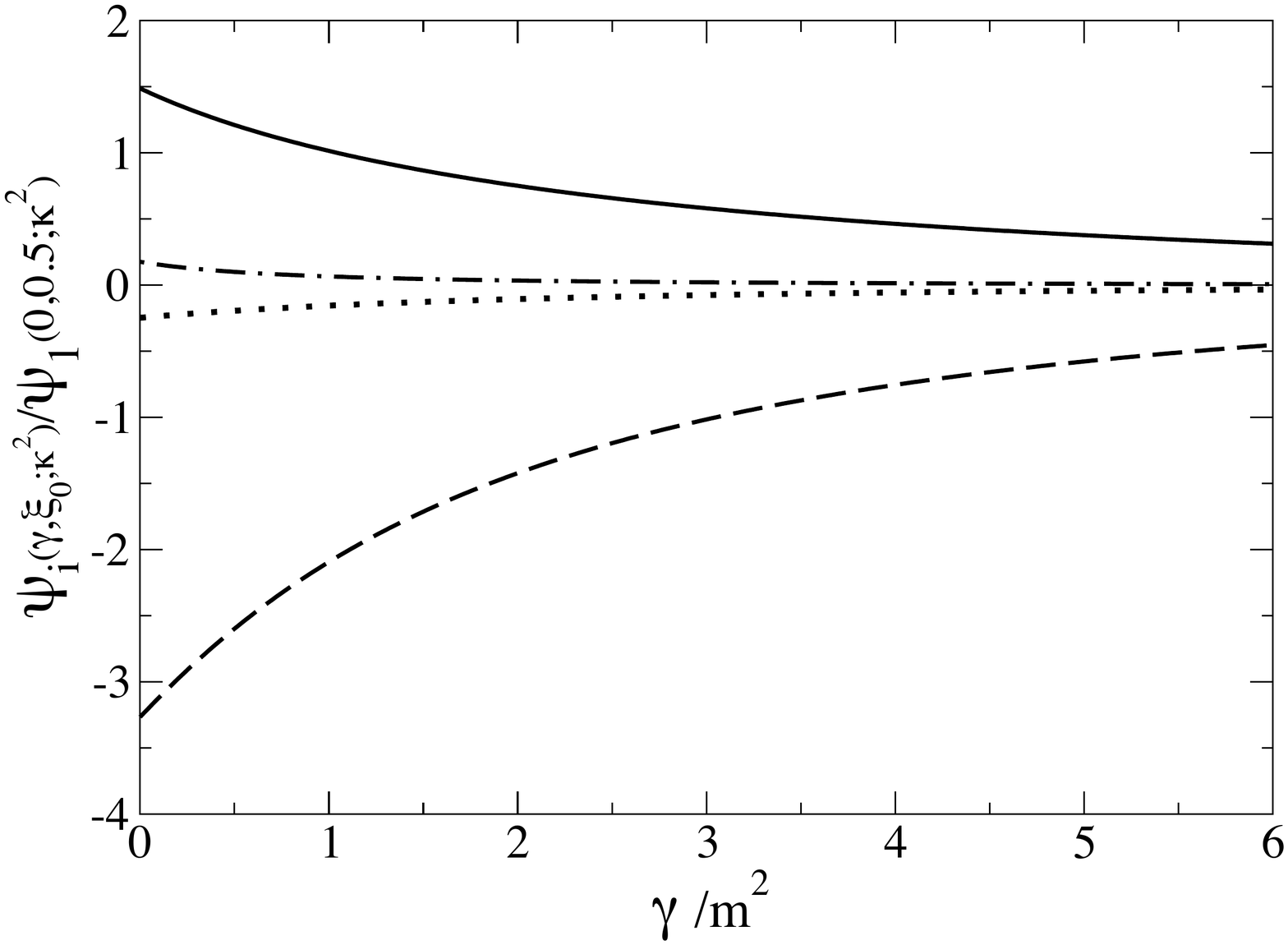}
    \includegraphics[width=8.0cm]{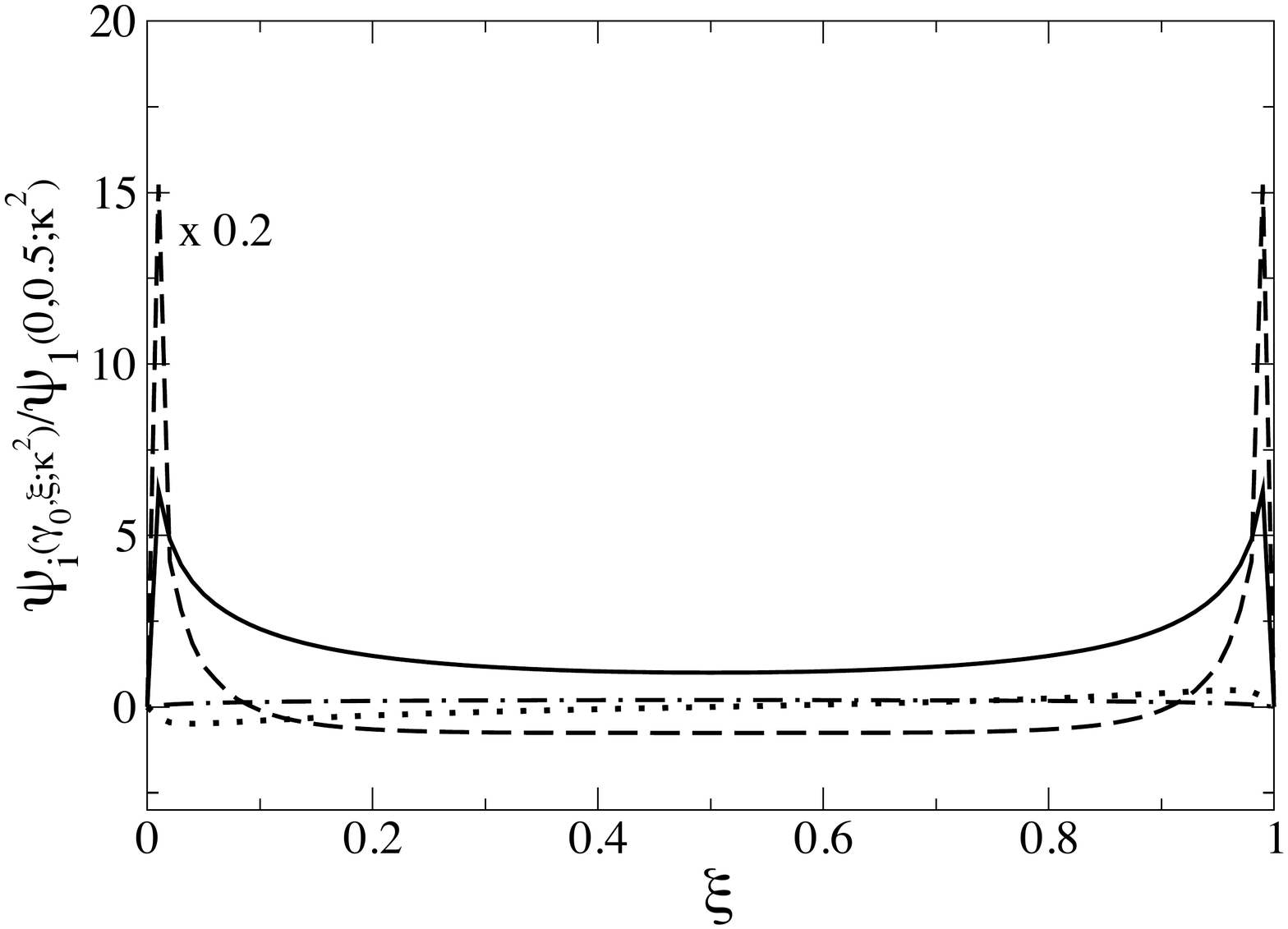}
    
    \vspace*{-0.5cm}
    \includegraphics[width=8.0cm]{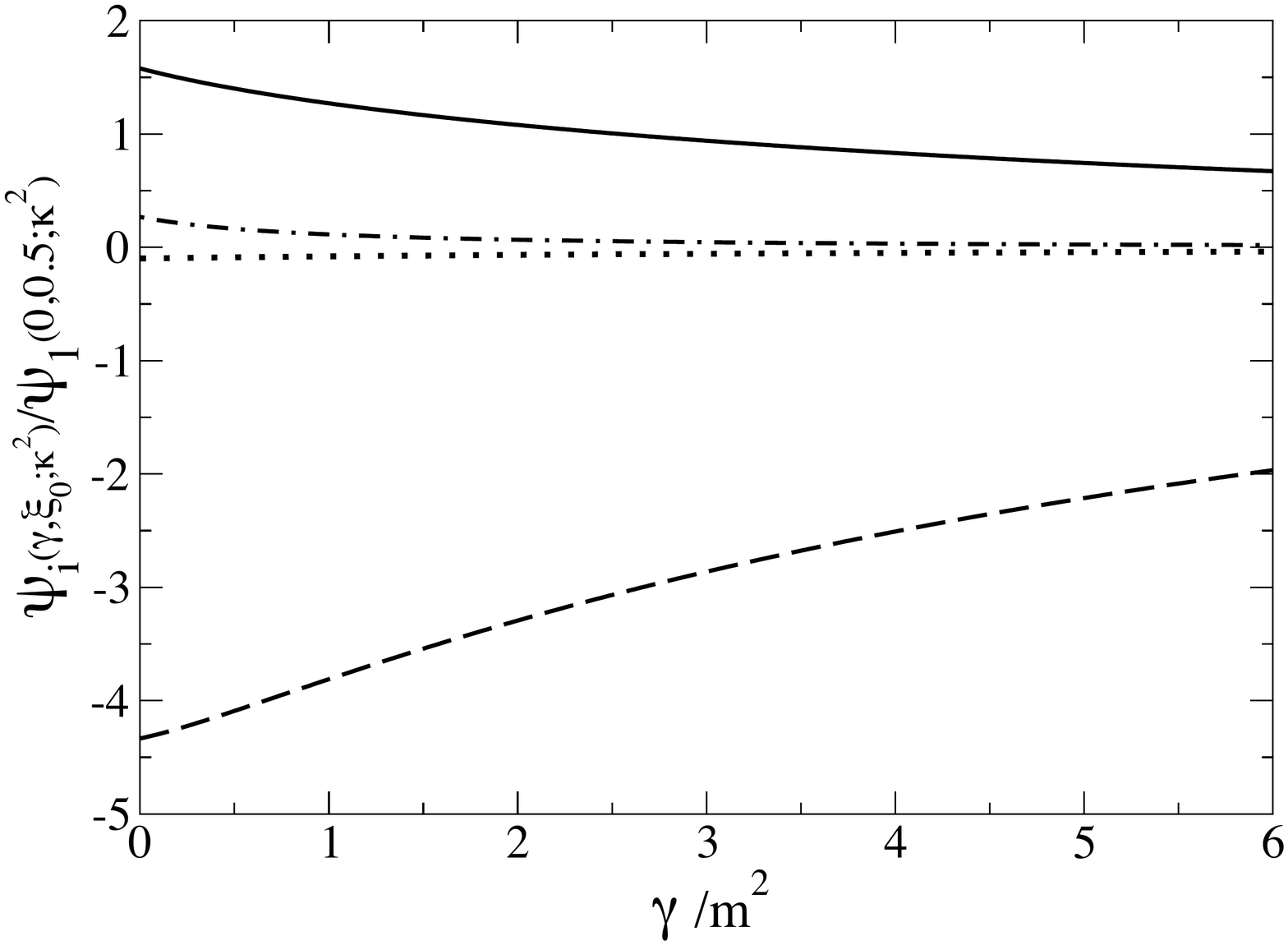}
    \includegraphics[width=8.0cm]{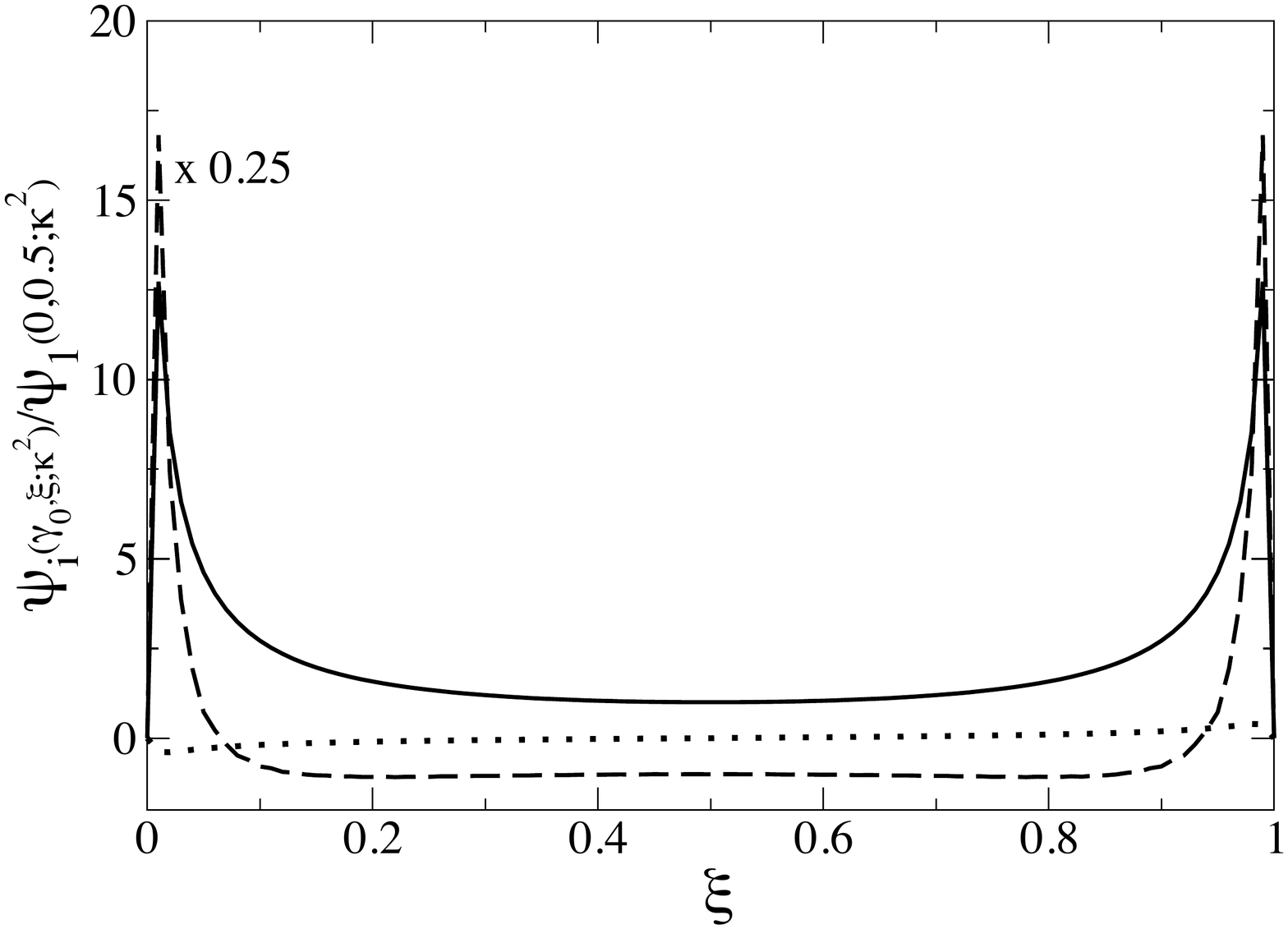}
   \caption{Light-front amplitudes  $\psi_i(\gamma,\,\xi;\kappa^2)$,  
   Eq. (\ref{psii}), for the pion-like system with a heavy-vector exchange 
$(\mu/m=2)$, 
    binding energy of $B/m=1.44$ and constituent mass $m=250$ MeV. Upper  
    panel:  vertex form-factor 
    cutoff $\Lambda/m=3$ and $g^2=435.0$, corresponding to $\alpha_s=10.68$ (see
    text for the definition of $\alpha_s$). 
    Lower  panel:
    vertex form-factor 
    cutoff $\Lambda/m=8$ and $g^2=53.0$, corresponding to $\alpha_s=3.71$.
    The value 
    of
    the longitudinal variable is $\xi_0=0.2$ and $\gamma_0=0$  .
     Solid line: $\psi_1$ . Dashed line:  $\psi_2$. Dotted line: $\psi_3$. Dot-dashed line: $\psi_4$}
  \label{fig:psipionl}
\end{figure*}
Lattice-QCD results obtained in Refs \cite{OLJPG11,DuaPRD16,OliFBS17}, by
adopting  the Landau gauge,  indicate 
that the gluon propagator obtained in the
space-like region is infrared finite and  can be
  roughly   described by using a massive propagator with a dynamical 
gluon mass $\mu \sim 500$ MeV. 
For the moment, we assume such a  { handy}
approximation for the gluon propagator  which  is able to qualitatively 
incorporate   
 the relevant momentum scale, but  in the Feynman gauge 
 (see Eq. (\ref{kernv})). Moreover,  Ref. \cite{ParPRD06} suggests 
for the lattice 
quark propagator  a
dynamical mass $m\sim 250$ MeV,
at zero momentum.  In conclusion, we  have adopted
i) a pion mass of about 140 MeV  leading to
   a binding energy $B/m=1.44$ (recall that $M=2m-B$), ii) a mass for the exchanged vector
   $\mu/m=2$, and iii) a cutoff in the vertex form factor $\Lambda/m$. For such
   a quantity, we have chosen two values,
$\Lambda/m=3$ and $\Lambda/m=8$, that correspond to  a size of the
interaction vertex less than the range of the interaction itself, 
and, according to Fig. 5.3 of Ref. \cite{Deur:2016tte},  yield a reasonable 
 {\em rescaled} coupling constant in the infrared region.  What we call  {\em rescaled} coupling constant, to stress the difference with $g^2$,
is defined 
by  (cf the vertex form factors in Eqs.  (\ref{LijNS}) and
(\ref{LijS}) as well as Eq. (\ref{bsefin}))
 \be \alpha_s={g^2\over 4\pi}~\left(1-{\mu^2\over \Lambda^2}\right)^2~~.
 \label{alphas}
 \ee
 Finally, it is worth noticing that  the assumed value of the constituent mass gives a  binding energy $B=1.44~m=360 $ MeV, that   
  leads to a typical size 
of $\hbar\,c /B=$ 0.55 fm, quite close to the experimental pion charge 
radius of 0.672$\pm$0.008 fm \cite{pdg}.

  The eigenvalues  $g^2$,
 obtained with the above set of parameters, and the corresponding $\alpha_s$ are
 summarized in Table \ref{tab2}, while the LF
   amplitudes of our {\em mock pion},  are presented in Fig. 
   \ref{fig:psipionl}.
 It is possible to recognize an overall  behavior of the $\psi_i$'s similar to the one  
 in Fig. \ref{fig:psivecgm0p15}, where a strong-binding case $B/m=1$ for a vector exchange
    with $\mu/m=0.15$ and $\Lambda/m=2$ is shown.  But it is much more interesting to observe 
    the effects produced by varying exchanged mass $\mu/m$ and vertex cutoff
     $\Lambda/m$. As  
   $\mu/m$  increases the tails of the LF amplitudes do the same (cf lhs of Figs.  \ref{fig:psivecgm0p15} 
   and \ref{fig:psipionl}): given the overall size of the system 
   (fixed by the binding energy), one has a
   reduction  of  the interaction range, that  in turn triggers a shrinking of the system itself. 
   The same kind of effect can be seen also decreasing the size of the interaction 
   vertex, i.e. by increasing $\Lambda/m$.
   More striking  is the effect shown in the rhs of Fig. \ref{fig:psipionl}, where the LF amplitudes
    are given as a function of the variable $\xi$. There, one can observe a quite
   peculiar  two-horn pattern, that becomes more enhanced when $\Lambda/m$ grows.
The striking two-horn pattern are  due to the 
presence of spin
degrees of freedom (see, the differences with the case of a two-scalar system in 
Ref. \cite{FSV2}).  Bearing 
 in mind that  we will  investigate  such structures in a forthcoming
  paper \cite{dFSV2},  we expect non trivial impacts of the above feature
on the evaluation of  both  
TMDs (see e.g. \cite{lorce2015unpolarized})  and valence  distributions (that contain  
 a multiplicative factor $\xi(1-\xi)$). In view of this, it  is suggestive to recall a well-known 
 pion distribution amplitude introduced in Ref. \cite{Chernyak:1981zz} 
 that displays 
a two-horn pattern.

\section{Conclusions and summary}\label{sect:conclusion}
  In the present paper, we have described in great detail our approach for getting
actual solutions, directly in Minkowski space, of the ladder Bethe-Salpeter equation for a system of two
 interacting fermions, in a state $0^+$, as well as a fermion\hyp{}antifermion 
 system in $0^-$ channel. This effort makes complete the presentation of our
  approach, shortly
 illustrated in Ref. \cite{dFSV1},  where the most urgent aim was  the
  quantitative comparison with
 the eigenvalues obtained in Ref. \cite{CK2010} and in the Euclidean space
 \cite{Dork}.   
The large amount  of provided information allows one  to assess in depth 
 the approach
based on the Nakanishi integral representation of the BS amplitude. Hence, one can easily   recognize that the 
approach is     
a very effective tool for exploring  the non perturbative  
dynamics inside a relativistic system, since it combines  flexibility (one is able 
to  address
higher-spin system) and  feasibility of affordable calculations.

The main ingredients for proceeding toward the numerical solution of  the coupled system of 
integral equations, formally equivalent to the initial BSE, are i) the
Nakanishi integral representation of the BS amplitude and ii) the LF-projection of the BS amplitude, that allows one to get LF amplitudes depending upon real variables.
 In order to embed  the approach based on NIR into a well-defined mathematical framework, 
 it is  fundamental  to notice that those LF distributions have 
exactly the form of a general  Stieltjes transform \cite{Carbonell:2017kqa}, and therefore the NIR approach
appears to be more general than one could suspect from the Nakanishi work focused on the diagrammatic
analysis of the transition amplitudes \cite{nak63,nak71}. 
The analysis of  the Stieltjes transform method 
for investigating the fermionic BSE will be left to future work.
Moreover, the LF projection plays also a key-role in determining the LF 
singularities that, in principle, plague the method (as shown 
in Ref. \cite{CK2010}, where a 
function was introduced for smoothing the singular behavior of the involved kernel). In particular, by exploiting previous analyses of the LF singularities \cite{Yan},
we succeeded in formally treating them. As a consequence, we  have obtained calculable matrix elements, after  introducing an orthonormal basis for expanding the four Nakanishi weight functions
needed for achieving the description of the fermionic states (for a $1^+$ states
the weights become eight).

 We have shown
 and discussed { both eigenvalues and eigenvectors  of the  generalized
 eigenvalue problem one gets 
 after
 inserting in the  BSE}
three different  kernels, featuring massive scalar, pseudoscalar and vector 
exchanges. Such interactions produce
very peculiar form for the correlation between the binding energy 
and the coupling constant $g^2$. { From the eigenvectors, i.e. the Nakanishi weight-functions, we have  evaluated the
corresponding LF amplitudes, that show the effects of the exchange in action. It
should be recalled }that the LF amplitudes  are the building-blocks for constructing 
transverse-momentum distributions
and valence wave functions (see, e.g. \cite{FSV1,FSV2}).
 Last but not least,
we have presented the application to a simplified model of the pion, by taking the mass of the constituents
and the mass of the exchanged vector boson (with a Feynman\hyp{}gauge propagator) 
from some typical lattice calculations.
 The intriguing feature shown in the rhs of Fig.
\ref{fig:psipionl} will be furtherly discussed in Ref. \cite{dFSV2}, 
where also the transverse\hyp{}momentum
distributions of a $0^-$ state will be investigated. 

In conclusion, we would like to emphasize that the present approach can be certainly enriched with new features impacting
both kernel and self-energies of both constituents and intermediate boson, but already
at the present  exploratory stage it appears to have a great potentiality, as  a
 phenomenological tool for facing with strongly relativistic systems, 
  { in  Minkowski space}.

\begin{acknowledgements}
We gratefully thank  J.
Carbonell and V. Karmanov for very stimulating  discussions.
TF and WdP acknowledge the warm hospitality of INFN Sezione di Roma and  thank 
 the partial financial 
support from the Brazilian Institutions:
 CNPq, CAPES and 
FAPESP. 
GS  thanks the partial support of CAPES
 and  acknowledges the warm hospitality of  the  Instituto Tecnol\'ogico de 
 Aeron\'autica.
\end{acknowledgements}
\newpage
\appendix

 \section{The coefficients $c_{ij}$ } \label{coeff}

In this Appendix,  the coefficients $c_{ij}$ in Eq. (\ref{coupls}) that determine the
kernel of the BSE  for scalar, pseudoscalar and vector exchanges, are briefly discussed. 
For the scalar exchange, one has

\begin{small}
\be
	c^S_{11} = m^2 +\frac{M^2}{4}-k^2, \,
	c^S_{12} = mM,\,
	c^S_{14} = -{B' \over M^2},\nonu
	c^S_{21} = m M, ~~
	c^S_{22} = m^2 +\frac{M^2}{4} +k^2 -2 \frac{(p \cdot k)^2}{M^2},\nonu
	c^S_{23} = -2 {B' \over M^4}(p \cdot k),\,
	c^S_{24} = -2 B'{m\over M^3}, 
	\nonu
	c^S_{32} = 2(p \cdot k), 
	c^S_{33} = \frac{B'}{B} \left( m^2 -\frac{M^2}{4} + 2 \frac{(p \cdot k)^2}{M^2} -k^2 \right), \,
	\nonu
	c^S_{34} = 2 \frac{B'}{B} \frac{m}{M} (p \cdot k), 
	c^S_{41} = M^2, ~~
	c^S_{42} = 2 m M, \nonu
	c^S_{43} = 2 \frac{B'}{B} \frac{m}{M} (p \cdot k), ~~
	c^S_{44} = -\frac{B'}{B} \left( \frac{M^2}{4} -m^2 - k^2 \right), \nonu \label{cij-1}
\ee
\end{small}
with $ c^S_{13} =c^S_{31} = 0$, $B$ and $B'$ defined by
\be
	B =  (p \cdot k)^2 - M^2 k^2\, \nonu
	B' = (p \cdot k)(p \cdot k'') - M^2 (k \cdot k'') ~ . \label{cij-2}
\ee
The other two sets of coefficients can be obtained from $c^S_{ij}$ by exploiting
the properties of the Dirac matrices. In particular, one gets for the
pseudoscalar exchange
\be
c_{i1}^{PS} = - c_{i1}^{S},  \quad c_{i2}^{PS} = c_{i2}^{S}, 
\quad c_{i3}^{PS} = c_{i3}^{S}, \nonu c_{i4}^{PS} = - c_{i4}^{S},~\forall i ~. \label{cijpseudo}\nonumber\\
\ee
and for 
the vector exchange
\be
c_{i1}^{V} = 4 ~ c_{i1}^{S},  \quad c_{i2}^{V} = -2 ~ c_{i2}^{S},
 \quad c_{i3}^{V} = -2 ~ c_{i3}^{S}, \nonu c_{i4}^{V} = 0,~\forall i ~. \label{cijvector}\nonumber\\
~~\ee

For carrying out the discussion of the LF singularities one meets, it is useful to
decompose the coefficients $c_{ij}(k,k'',p)$ so that terms with the same power
of $k^\mu$ are gathered together. Namely
\be
c^S_{ij}(k,k'',p) = a^0_{ij}+ a^1_{ij}~(p\cdot k)+a^2_{ij}~(p\cdot k)^2
\nonu +a^3_{ij}~k^2 
+ {1 \over B}\Bigl[(p\cdot k) (p\cdot k'')-M^2
(k\cdot k'')\Bigr]  \nonu 
\times \Bigl[b^0_{ij}+ b^1_{ij}~( p\cdot k) + b^2_{ij}
~(p\cdot k)^2+b^3_{ij} ~k^2\Bigr]
 \nonu
+ \Bigl[(p\cdot k) (p\cdot k'')-M^2
(k\cdot k'')\Bigr]
\nonu \times~\Bigl[d^0_{ij}+d^1_{i,j}~(p\cdot k) \Bigr],
\label{coef}\ee 
where the non vanishing coefficients $a^{n}_{ij}$, $b^{n}_{ij}$ and $d^{n}_{ij}$ are given by

\begin{small} 
\be
a_{11}^0 = m^2+M^2/4 \, ,
a_{12}^0 =  mM \, ,
a_{21}^0 =  mM \, ,\nonu
a_{22}^0 =m^2+M^2/4 \, ,
a_{41}^0 =  M^2 \, ,
a_{42}^0 =  2mM \, ,
a_{32}^1 =  2 \, ,\nonu
a_{22}^2 = - 2/M^2 \, ,
a_{11}^3 =  -1 \, ,
a_{22}^3 =  1 \, ,\nonu
b_{33}^0 =  m^2-M^2/4 \, ,
b_{34}^1 = 2m/M\, ,
b_{43}^1 = 2m/M\, ,\nonu
b_{44}^1 = m^2-M^2/4\, ,
b_{33}^2 = 2/M^2\, ,
b_{33}^3 = -1\, , \nonu
d_{14}^0 = -1/M^2\, ,
d_{24}^0 = -2m/M^3\, ,
d_{44}^0 = 1\, ,
d_{23}^1 = -2/M^4\, .
\nonu\label{tablecoeffaijk}
\ee 
\end{small}

\section{Analytic integration on $k^-$ of the kernel ${\cal L}_{ij}$ for any
$k^+_D$ 
 } \label{genint} 

This Appendix contains the details on the  integration 
in the complex $k^{-}$-plane of the
kernel ${\cal L}_{ij}$, Eq. (\ref{Lij1}), for any value of $k^+_D$.  Namely, we obtain the general expression of the 
coefficients ${\cal C}_n$ defined in Eq. (\ref{cn}). They are necessary for
determining the kernel ${\cal L}_{ij}$.

One can write
\be
{\cal C}_n =
~- {\partial \over \partial \ell_D}
{\cal B}_n
\ee
where $n=0,1,2,3$ and

\begin{small}
\be
{\cal B}_n
=
2\int {dk^-\over 2 \pi}~{ (k^-)^n ~
\over
 \Bigl[(1-z)(k^- - k^-_d)+i\epsilon\Bigr]}\nonu \times~
{1\over \Bigl[(1+z)(k^- -k^-_u)-i\epsilon\Bigr]}
\nonu
\times\int^1_0 d\xi_1
~ {\xi_1 \over \Bigl[k^+_D k^-+\ell_D 
+\xi_1 (1-v)\Bigl ( \mu^2 -\Lambda^2\Bigr) +
i\epsilon\Bigr]^3}
\nonu\ee
\end{small}
where 
it has been applied the Feynman trick to the denominators containing $k^+_D$ 
for obtaining the final expression.

 The issue is the possibility or not to apply the Cauchy's
residue theorem for getting an analytic result. Hence, one has to carefully
check if the arc at infinity gives or not a vanishing contribution. As it is
shown in what follows, a positive answer depends upon the values of $n$ and
$k^+_D$.

The $n=0$ case is not affected by any difficulty due to values of $k^+_D$. 
One can always close the arc at infinity, even for $k^+_D=0$.
Indeed, if the last case happens, one could be concerned about the end-points
  $z= \pm 1$,  but, fortunately, 
one remains with a standard LF integration and gets a finite result 
(see Ref. \cite{bakker2005restoring}). The evaluation of ${\cal B}_0$ proceeds
 by exploiting the residue theorem, obtaining
\be
{\cal B}_0 =
{i\Bigl[\theta (k^+_D)~{\cal I}( k^-_u)
+\theta (-k^+_D)~{\cal I}( k^-_d)\Bigl] \over (1-z^2)~(k^-_u-k^-_d)}~,
\ee
where

\begin{small}
\be
{\cal I}( k^-_u) = 
 {1 \over \Bigl[k^+_D k^-_u+\ell_D 
+ (1-v)\Bigl ( \mu^2 -\Lambda^2\Bigr) +
i\epsilon\Bigr]^2} \nonu
\times ~{1\over \Bigl[k^+_D k^-_u+\ell_D  +
i\epsilon\Bigr]}
\ee
\end{small}
and the analogous expression for ${\cal I}( k^-_d)$.
Then ${\cal C}_0$ is given by

\begin{small}
\be
{\cal C}_0 = -{i \over (1-z^2)~(k^-_u-k^-_d)}
\nonu \times~\left[\theta (k^+_D)~{\partial \over \partial \ell_D}
{\cal I}( k^-_u)+
+\theta(-k^+_D)~{\partial \over \partial \ell_D}
{\cal I}( k^-_d)\right]~.
\ee
\end{small}

For $n=1$, the integral ${\cal B}_1$ is
\be
{\cal B}_1 = I_1+ k^-_u {\cal B}_0
\ee
with
\be
I_1 =
-{2i\over (1-z^2)} \theta(-k^+_D)
~
\int_0^1d\xi_1~\xi_1
\nonu \times~{1_1\over 
\Bigl\{k^+_D k^-_d+\ell_D 
+\xi_1 (1-v)\Bigl ( \mu^2 -\Lambda^2\Bigr) +i\epsilon\Bigr\}^3} ~.  \nonu
\ee
Let us remind that $-(1-z)M/(k^+_D-(1-z)M)\ge 0$ for $0\ge k^+_D$.

Summarizing, one gets
\be
{\cal B}_1 =
{i\Bigl[k^-_u 
\theta(k^+_D)~ {\cal I}(k^-_u)+k^-_d\theta(-k^+_D)~ {\cal I}(k^-_d)\Bigr]
\over (1-z^2) (k^-_u-k^-_d)}~ .
\nonu \ee

  Then ${\cal C}_1$ is given by
  
 \begin{small}
 \be
 {\cal C}_1= -{i\over (1-z^2) (k^-_u-k^-_d) } \nonu \times~
 \left[k^-_u~\theta(k^+_D)~{\partial\over \partial \ell_D} {\cal
 I}(k^-_u) +
 k^-_d~\theta(-k^+_D)~{\partial\over \partial \ell_D} {\cal
 I}(k^-_d)\right]~.\nonu
 \ee
 \end{small}

For $n=2$, the integral ${\cal B}_2$ is

 \be
{\cal B}_2 
= {i \,\delta(k^+_D)\over (1-z^2) \Bigl[ (1-v)\Bigl ( \mu^2 -\Lambda^2\Bigr)
\Bigr]^2}\nonu \times ~
\left[{(1-v)\Bigl ( \mu^2 -\Lambda^2\Bigr)\over \Bigl[\ell_D 
+ (1-v)\Bigl ( \mu^2 -\Lambda^2\Bigr) +
i\epsilon\Bigr]}  - L(\ell_D)
\right]
\nonu+ {i\Bigl[(k^-_u)^2 
\theta(k^+_D)~ {\cal I}(k^-_u)+(k^-_d)^2\theta(-k^+_D)~ {\cal I}(k^-_d)\Bigr]
\over (1-z^2) (k^-_u-k^-_d)}~.  \nonu  
\ee
where
\be
L(\ell_D)=\ln\left({\ell_D 
+ (1-v)\Bigl ( \mu^2 -\Lambda^2\Bigr)\over 
\ell_D}\right)
\ee

Then, ${\cal C}_2$ is given by the sum of a singular term and the non singular
one already obtained, i.e.
\be
{\cal C}_2=
-{i \over (1-z^2)}
~{\delta(k^+_D)~\over \ell_D
\Bigl[\ell_D 
+ (1-v)\Bigl ( \mu^2 -\Lambda^2\Bigr) +
i\epsilon\Bigr]^2}~
\nonu
-{i\over (1-z^2) (k^-_u-k^-_d)}~ 
\Bigl[(k^-_u)^2 
\theta(k^+_D)~  {\partial \over \partial \ell_D}{\cal I}(k^-_u) \nonu
+ (k^-_d)^2\theta(-k^+_D)~  {\partial \over \partial \ell_D}{\cal I}(k^-_d)\Bigr]
\ee

For $n=3$, the integral ${\cal B}_3$ is

\be
{\cal B}_3=
~{i \over (1-z^2)}~{\partial \over \partial k^+_D}
\delta(k^+_D)~{1 \over \Bigl[(1-v)\Bigl ( \mu^2 -\Lambda^2\Bigr)\Bigr]^2}
\nonu \times ~
\left[(1-v)\Bigl ( \mu^2 -\Lambda^2\Bigr) 
- \ell_D ~L(\ell_D)
\right]\nonu
+(k^-_u+k^-_d)~{\cal B}_2 - k^-_u k^-_d~{\cal B}_1
\nonu
\ee

Then ${\cal C}_3$ contains both singular and non singular contributions as in the
case of ${\cal C}_2$. One has 
\be
{\cal C}_3 =
{i \over (1-z^2)}~{\partial \over \partial k^+_D}
\delta(k^+_D)~{1 \over \Bigl[(1-v)\Bigl ( \mu^2 -\Lambda^2\Bigr)\Bigr]^2}
\nonu \times~\left[ L(\ell_D) 
 -{{ (1-v) }  (\mu^2-\Lambda^2)
\over \ell_D +{ (1-v) }  (\mu^2-\Lambda^2)}\right]
\nonu
-{i~(k^-_u+k^-_d) \over (1-z^2)}
~{\delta(k^+_D)~\over \ell_D
\Bigl[\ell_D 
+ (1-v)\Bigl ( \mu^2 -\Lambda^2\Bigr) +
i\epsilon\Bigr]^2}~
\nonu
-{i\over (1-z^2) (k^-_u-k^-_d)}~ 
\Bigl[(k^-_u)^3 
\theta(k^+_D)~  {\partial \over \partial \ell_D}{\cal I}(k^-_u)\nonu
+
(k^-_d)^3\theta(-k^+_D)~  {\partial \over \partial \ell_D}{\cal I}(k^-_d)\Bigr]
\ee

\section{Singular terms for scalar, pseudoscalar and vector interactions}
\label{singterm}
In this Appendix,  
the final expressions of the singular contributions 
 obtained in the previous  \ref{genint} are explicitly given, in order to
 facilitate a quick  usage of the main results of our work. For the scalar case, the non vanishing contributions for the singular part 
of ${\cal L}^{(s)}_{ij}$   are 

\begin{small}
\be
{\cal L}^{(s)}_{14} ={i\over M} ~{(\mu^2-\Lambda^2)^2\over 8\pi^2 M^2}~
{\delta(z'-z) \over 2 (1-z^2)}
\int^1_0 dv~{v~(1-v)^2\over D_\ell}
 \nonu
{\cal L}^{(s)}_{22} ={i\over M} ~{(\mu^2-\Lambda^2)^2\over 8\pi^2  M^2}~
{\delta(z'-z) \over  (1-z^2)}
\int^1_0 dv~{v~(1-v)\over D_\ell}
 \nonu
{\cal L}^{(s)}_{24} ={i\over M} ~{(\mu^2-\Lambda^2)^2\over 8\pi^2 M^2}~{m\over M}~
{\delta(z'-z) \over  (1-z^2)}
\int^1_0 dv~{v~(1-v)^2 \over D_\ell}
 \nonu
{\cal L}^{(s)}_{33} =-{i\over M} ~{(\mu^2-\Lambda^2)^2\over 8\pi^2 M^2}~
{\delta(z'-z) \over  (1-z^2)}
\int^1_0 dv~{v~(1-v)^2\over D_\ell}\nonu
{\cal L}^{(s)}_{23}  = {\cal L}^{(s)}_{23}(a) + {\cal L}^{(s)}_{23}(b)
\ee
\end{small}
with

\begin{small}
 \be
{\cal L}^{(s)}_{23}(a) ={i\over M} ~{(\mu^2-\Lambda^2)^2\over 8\pi^2 M^2}~\delta(z'-z)~{2z \over M^2 (1-z^2)^2}\nonu
\times
\Bigl[{M^2~(1-z^2)\over 8}  
+  \gamma +m^2 \Bigr]\int^1_0 dv~{v~(1-v)^2
\over D_\ell } 
\ee

\be
 {\cal L}^{(s)}_{23}(b) ={i\over M} ~{1\over 8\pi^2 M^4~(1-z^2)} 
 ~\Bigl[{\partial \over \partial z'}\delta(z'-z)\Bigr]
 \nonu\times\int^1_0 dv~
{1\over (1-v)} ~
 \left[{(1-v)\Bigl ( \mu^2 -\Lambda^2\Bigr)
\over  \ell_D
+(1-v)\Bigl ( \mu^2 -\Lambda^2\Bigr) 
} -L(\ell_D)  \right]\nonu
\label{l23sab}
\ee 
\end{small}
In the above expressions

\begin{small}
\be
D_{\ell}=\tilde \ell_D
\Bigl[\tilde \ell_D 
+ (1-v)\Bigl ( \mu^2 -\Lambda^2\Bigr) +
i\epsilon\Bigr]^2
\nonu 
\tilde \ell_D=-v(1-v) ~\gamma
-v\Bigl(\gamma'  +z^2m^2+ (1-z^2)\kappa^2\Bigr) 
\nonu-(1-v)\mu^2
\ee
\end{small}
Notice that even for $\Lambda^2<\mu^2$ the factor $D_\ell$ does not vanish, due to the factor $(1-v)\mu^2$ in $\tilde{\ell}_D$ which is canceled by a corresponding 
factor in the square bracket of $D_\ell$. Finally, the singular terms for pseudoscalar $({\cal L}_{ij}^{(s)})^{PS}$ and 
vector exchange $({\cal L}_{ij}^{(s)})^{V} $ can be written in terms of the singular terms for a 
scalar boson exchange written  above for ${\cal L}_{ij}^{(s)}$. From Eqs. (\ref{cijpseudo}) and (\ref{cijvector}) one gets for the pseudoscalar exchange

\begin{small}
\be
({\cal L}_{14}^{(s)})^{PS} = - {\cal L}_{14}^{(s)} ~, \quad 
({\cal L}_{22}^{(s)})^{PS} = {\cal L}_{22}^{(s)} ~, \nonu ({\cal L}_{24}^{(s)})^{PS} = - {\cal L}_{24}^{(s)} ~, 
 ({\cal L}_{33}^{(s)})^{PS} = {\cal L}_{33}^{(s)} ~, \nonu ({\cal L}_{23}^{(s)})^{PS} = {\cal L}_{23}^{(s)} ~.\nonu
\ee
\end{small}
while for 
the vector exchange one has

\begin{small}
\be
({\cal L}_{14}^{(s)})^{V} = 0 ~, \quad ({\cal L}_{22}^{(s)})^{V} = -2 ~{\cal L}_{22}^{(s)} ~, \quad ({\cal L}_{24}^{(s)})^{V} = 0 ~, \nonu
 ({\cal L}_{33}^{(s)})^{V} = -2 ~ {\cal L}_{33}^{(s)} ~,  \quad ({\cal L}_{23}^{(s)})^{V} = -2~{\cal L}_{23}^{(s)} ~.\nonu
\ee
\end{small}
\bibliographystyle{epjc}
\bibliography{naka}

\end{document}